\documentclass[twocolumn]{aastex63}
\usepackage{natbib}
\usepackage{amsmath}
\usepackage{lineno}
\usepackage{multirow}
\defcitealias{yuan19}{Y19}

\def\deg{^\circ}
 
\def\kpc{{\rm\,kpc}}
\def\kms{{\rm\,km\,s^{-1}}}
\def\msun{{\rm\,M_\odot}}

\def\pc{{\rm\,pc}}

\def\lsun{{\rm L}_\odot}

\def\eg{{ e.g.,\ }}
\def\ie{{ i.e.,\ }}
\def\lta{\mathrel{\spose{\lower 3pt\hbox{$\mathchar"218$}}
     \raise 2.0pt\hbox{$\mathchar"13C$}}}
\def\gta{\mathrel{\spose{\lower 3pt\hbox{$\mathchar"218$}}
     \raise 2.0pt\hbox{$\mathchar"13E$}}}

\def\FeH{{\rm[Fe/H]}}

\submitjournal{ApJ}

\shorttitle{Cetus from StreamFinder and StarGO}
\shortauthors{Yuan et al.}
\begin{document}


\title{The Complexity of the Cetus Stream Unveiled from the Fusion of STREAMFINDER and StarGO}

\author[0000-0002-8129-5415]{Zhen Yuan}
    \affiliation{Universit\'e de Strasbourg, CNRS, Observatoire Astronomique de Strasbourg, UMR 7550, F-67000 Strasbourg, France}
    \email{zhen.yuan@astro.unistra.fr}

\author{Khyati Malhan}
\affiliation{Max-Planck-Institut f\"ur Astronomie, K\"onigstuhl 17, D-69117, Heidelberg, Germany}

\author{Federico Sestito}
\affiliation{Department of Physics and Astronomy, University of Victoria, Victoria, BC V8W 3P2, Canada}

\author{Rodrigo A. Ibata}
\affiliation{Universit\'e de Strasbourg, CNRS, Observatoire Astronomique de Strasbourg, UMR 7550, F-67000 Strasbourg, France}

\author{Nicolas F. Martin}
\affiliation{Universit\'e de Strasbourg, CNRS, Observatoire Astronomique de Strasbourg, UMR 7550, F-67000 Strasbourg, France}
\affiliation{Max-Planck-Institut f\"ur Astronomie, K\"onigstuhl 17, D-69117, Heidelberg, Germany}

\author{Jiang Chang}
\affiliation{Purple Mountain Observatory, CAS, No.10 Yuanhua Road, Qixia District, Nanjing 210034, China}

\author{Ting S. Li}
\affiliation{Department of Astronomy and Astrophysics, University of Toronto, 50 St. George Street, Toronto ON, M5S 3H4, Canada}

\author{Elisabetta Caffau}
\affiliation{GEPI,Observatoire de Paris, Universit\'e PSL, CNRS, 5 Place Jules Janssen, 92190 Meudon, France}

\author{Piercarlo Bonifacio}
\affiliation{GEPI,Observatoire de Paris, Universit\'e PSL, CNRS, 5 Place Jules Janssen, 92190 Meudon, France}

\author{Michele Bellazzini}
\affiliation{INAF - Osservatorio  di  Astrofisica  e  Scienza  dello  Spazio,via Gobetti 93/3, 40129 Bologna, Italy}

\author{Yang Huang}
\affiliation{South-Western Institute for Astronomy Research, Yunnan University, Kunming 650500, People’s Republic of China}

\author{Karina Voggel}
    \affiliation{Universit\'e de Strasbourg, CNRS, Observatoire Astronomique de Strasbourg, UMR 7550, F-67000 Strasbourg, France}
    
\author{Nicolas Longeard}
\affiliation{Laboratoire d’astrophysique, \'Ecole Polytechnique F\'ed\'erale de Lausanne (EPFL), Observatoire, 1290 Versoix, Switzerland}

\author{Anke Arentsen}
    \affiliation{Universit\'e de Strasbourg, CNRS, Observatoire Astronomique de Strasbourg, UMR 7550, F-67000 Strasbourg, France}

\author{Amandine Doliva-Dolinsky}
    \affiliation{Universit\'e de Strasbourg, CNRS, Observatoire Astronomique de Strasbourg, UMR 7550, F-67000 Strasbourg, France}
    
\author{Julio Navarro}
\affiliation{Department of Physics and Astronomy, University of Victoria, Victoria, BC V8W 3P2, Canada}

\author{Benoit Famaey}
\affiliation{Universit\'e de Strasbourg, CNRS, Observatoire Astronomique de Strasbourg, UMR 7550, F-67000 Strasbourg, France}

\author{Else Starkenburg}
\affiliation{Kapteyn Astronomical Institute, University of Groningen, Landleven 12, NL-9747 AD Groningen, the Netherlands}

\author{David S. Aguado}
\affiliation{Dipartimento di Fisica e Astrofisica, Univerisit\'a degli Studi di Firenze, via G. Sansone 1, Sesto Fiorentino 50019, Italy}


\begin{abstract}
We combine the power of two stream-searching tools, \texttt{STREAMFINDER} and \texttt{StarGO} applied to the \emph{Gaia} EDR3 data, to detect stellar debris belonging to the Cetus stream system that forms a complex, nearly polar structure around the Milky Way. In this work, we find the southern extensions of the northern Cetus stream as the Palca stream and a new southern stream, which overlap on the sky but have different distances. These two stream wraps extend over more than $\sim100\deg$ on the sky ($-60\deg<\delta<+40\deg$). The current N-body model of the system reproduces both as two wraps in the trailing arm. We also show that the Cetus system is confidently associated with the Triangulum/Pisces, Willka Yaku, and the recently discovered C-20 streams. The association with the ATLAS-Aliqa Uma stream is much weaker. All of these stellar debris are very metal-poor, comparable to the average metallicity of the southern Cetus stream with $\FeH=-2.17\pm0.20$. The estimated stellar mass of the Cetus progenitor is at least $10^{5.6}\msun$, compatible with Ursa Minor or Draco dwarf galaxies. The associated globular cluster with similar stellar mass, NGC 5824 very possibly was accreted in the same group infall. The multi-wrap Cetus stream is a perfect example of a dwarf galaxy that has undergone several periods of stripping, leaving behind debris at multiple locations in the halo. The full characterization of such systems is crucial to unravel the assembly history of the Milky Way and, as importantly, to provide nearby fossils to study ancient low-mass dwarf galaxies.

\end{abstract}


\section{Introduction}
\label{sec:intro}

Low-mass dwarf galaxies are dark matter-dominated systems thought to play an important role in the hierarchical formation of galactic dark matter halos. The search for the most extreme examples of these systems in the vicinity of the Milky Way has been transformed by the systematic mapping of the sky made possible by large panoptic surveys. In particular, the Sloan Digital Sky Survey \citep[SDSS;][]{york00} led to the discovery of many new dwarf galaxies significantly fainter than previously known systems \citep[\eg][]{willman05a, willman05b, belokurov06, irwin07,koposov08}. This revolution, initiated with the SDSS, was continued through a series of similar surveys conducted in the last decade, including Pan-STARRS1 \citep[PS1;][]{chambers16} and the Dark Energy Survey \citep[DES;][]{des}. Trawling these data sets led to the discovery of tens of new faint dwarf galaxies that orbit the Milky Way \citep[\eg][and references within]{bechtol15, koposov15, laevens15b, martin15, wagner15}.

Low-luminosity dwarf galaxies, expected to inhabit the lowest-mass dark matter halos that can form stars \citep[e.g.][]{bullock17}, garner a lot of interest as they are thought to be direct fossils from the very early universe \citep{bovill09}. The faintest of all, with only 10$^3$ to $10^5\msun$ in stars, are observed to have short star formation histories, limited to the first 1--2 Gyr after the Big Bang ($z$ $\sim$ 5; e.g. \citealt{brown14}), and have average metallicities $\FeH\la-2$ \citep{simon19}. Because of their relatively simple chemical enrichment history, the elemental abundances of stars from these systems are extremely sensitive to star-forming activities, the initial mass function, and neutron capture events ($s$- and $r$-process). These information is decoded using high-resolution (HR) spectra of stars in these ancient systems. However, the closest dwarf satellites remain $\sim30\kpc$ away from us, which significantly limits the number of bright stars amenable to high signal-to-noise HR spectroscopic observations before the advent of 30-meter telescopes.


On the other hand, the hierarchical formation that every galaxy undergoes means that numerous low-mass dwarf galaxies were accreted onto the MW and, during their disruption, they left relics in the stellar halo \citep[see e.g.][]{johnston98,bullock05,amorisco17,monachesi19}. Identifying halo stars that previously belonged to the same low-mass dwarf galaxy would allow us to study their ancient progenitor with more detailed spectroscopic information than in current dwarf galaxies, as their stellar debris might extend to much closer distances. Thanks to the ESA/\emph{Gaia} mission \citep{gaia}, discovering the very low surface brightness remnants of tidally disrupted low-mass dwarf galaxies is now an achievable goal. In synergy with the efforts designed to search for low-metallicity stars \citep[\eg][]{beers05,starkenburg17,li18}, many studies have started to unveil the traces of low-metallicity accreted systems \citep[][]{sestito19, sestito20, wan20, yuan20b, yuan20a, martin22a, martin22b}. 

In parallel, the library of stellar debris is continuously increasing, including streams of globular cluster origins, detected based on their coherence in phase space \citep[see \eg][]{price18, malhan18b, bonaca19, ibata19a, ibata21, li21b}, and substructures of dwarf galaxy origins, identified from their clustering signatures in dynamical space \citep[see e.g.][]{helmi99, belokurov18, helmi18, koppelman19, myeong18a, myeong19, matsuno19}. A list of studies have shown that these disrupted stellar systems, together with globular clusters and chemically peculiar stars with halo orbits, can be grouped based on similar orbital properties, suggestive of their common origins \citep{roederer18, myeong18b, massari19, myeong19, naidu20, yuan20a, bonaca21, limberg21, gudin21, li21b, shank21}. Detailed abundance studies for these stellar debris have been made possible by HR-spectroscopic follow-up studies \citep{ji20,aguado21a, aguado21b, matsuno21, gull21}, as well as by HR-spectroscopic surveys, such as the GALactic Archaeology with HERMES survey \citep[GALAH;][]{buder21,simpson21}.

In this work, we focus on the systematic search for members of a disrupted low-mass dwarf galaxy, the Cetus stream system, and assess its associations with several stellar relics that possibly share a common origin. A tailored N-body model is also used to interpret its stripping history. This is one of the few dwarf galaxy streams that preserve coherent structures in configuration space and allow us to decode their disruption history through modeling. The other two such systems include the text-book example of the Sagittarius (Sgr) stream \citep{mateo96, ibata01, majewski03} with an extensive list of simulation studies \citep[see e.g.][]{penarrubia10, law10, vera13, dierickx17, vasiliev21b} and the LMS-1 from recent studies \citep{yuan20b,malhan21}. Ultimately, the list of confident Cetus members should yield a sizable sample of stars to study the chemical evolution of a small and ancient dwarf galaxy.

The Cetus stream was first discovered by \citet{newberg09} using data from the SDSS and they suggested an association with the globular cluster NGC5824 based on radial velocities. With the second data release of $Gaia$ \citep{brown18}, \citet[][hereafter, \citetalias{yuan19}]{yuan19} identified $\sim 150$ Cetus members by their clustering in kinematic space and confirmed the association with NGC5824 based on the similarity of their orbits. They showed that Cetus is comprised of two parts on opposite sides of the Galactic disc where it is difficult to track its stars. Both parts of the stream have a mean metallicity $\FeH\approx-$2 with an intrinsic metallicity dispersion of 0.1 -- 0.2 dex, suggesting its progenitor was a low-mass dwarf galaxy according to the stellar mass--metallicity relation \citep{kirby13}. Based on the more densely populated wrap identified in the Galactic South, \citet{chang20} performed an N-body model of the stream and predicted that about half of its members are distributed in the southern sky. In the model, the southern extension of the Cetus stream overlaps the location of the Palca stream discovered in the DES \citep{shipp18}, with a compatible distance. If this association is confirmed, it would extend the known length of Cetus by an additional $\sim40\deg$. The Southern Stellar Stream Spectroscopy Survey \citep[$S$5;][]{li19} reported obsrvations of 25 Palca members in the field of the ATLAS-Aliqa-Uma (AAU) stream \citep{li21a}, which overlaps Palca on the sky but with a perpendicular stream track. Most of the Palca members near AAU have kinematics consistent with the Cetus model \citep{li21b}, as will be shown in detail in this study. This confirmation encourages us to search for more Cetus-Palca members in the southern sky. 

To explore the southern hemisphere, where spectroscopic data is lacking, for extension of the Cetus stream we take advantage of the most up-to-date stream catalog derived with the \texttt{STREAMFINDER} algorithm \citep{malhan18a, ibata19a, ibata21} as it does not require radial velocity measurements. The algorithm evaluates the probabilities of stars being in streams based on the similarity of their orbital properties with those of their neighbors. \texttt{STREAMFINDER} has proven hugely successful, with numerous new discoveries \citep{malhan18b}, including the extended Phlegethon stream \citep{ibata18}, and the long sought-after stream of $\omega$-Cen \citep{ibata19b} predicted by simulations \citep[see \eg][]{bekki03, mizutani03, ideta04, tsuchiya04}. Using the newly released \emph{Gaia} EDR3 and continuous optimizations of the search algorithms, the Milky Way’s stream landscape has been recently updated \citep{ibata21}. The most recent \texttt{STREAMFINDER} catalog includes a search for wide streams, with which we detected the most nearby dwarf galaxy stream ($d\sim20\kpc$), LMS-1 \citep{malhan21}. These streams are highly coherent in kinematic space, and therefore clean stream tracks can be revealed by exerting cuts on proper motion and significance, the latter of which quantifies the stream-like behavior. 

In the case of the Cetus stream, it is known to be spatially separated, and diffusely distributed in phase space (\citetalias{yuan19}), which makes it difficult to implement clean membership filters with simple kinematic cuts. In order to select the most likely Cetus members, we fuse \texttt{STREAMFINDER} with another stream searching algorithm, \texttt{StarGO} (Stars' Galactic Origins), developed from a totally different perspective \citep{yuan18}. The latter algorithm is a neural-network based clustering tool, built on one of the most popular unsupervised learning algorithms, the self-organizing-map (SOM). Utilizing the power of SOM to store and visualize n-D data structures, a systematic group identification procedure was developed to search for streams and substructures clustered in dynamical space. The underlying assumption is that stars sharing the same origins preserve their clustering signatures in their orbital properties after they are stripped from their progenitor. \texttt{StarGO} has successfully led to the identification of the Cetus stream in \citetalias{yuan19}, the discovery of the LMS-1 structure \citep{yuan20b}, and a plethora of dynamically tagged groups (DTGs) in the nearby stellar halo \citep{yuan20a}. 

Our strategy in the present contribution is to first obtain a sample of $Gaia$ EDR3 stars that are likely to be in streams identified by \texttt{STREAMFINDER}, together with their orbital properties given their most likely orbits as derived by this algorithm. \texttt{STREAMFINDER} does not, however, link together stars that are part of the same stream. We therefore apply \texttt{StarGO} to the selected sample, and identify dynamically tagged groups (DTGs) that have similar properties with the known Cetus stream. In the previous applications of \texttt{StarGO}, the dynamical parameters were derived from observational quantities, whereas, here, they are calculated from the predicted values of \texttt{STREAMFINDER}. This fusion of the two methods allows us to get the most likely candidate member list for the Cetus stream in the southern sky, where line-of-sight velocity information is largely missing. Further confirmation of the membership requires radial velocity measurements, as addressed below.

We describe the detailed detection procedure in Sec.~\ref{sec:method}. After we get the new member list for the Cetus system, the data used for member confirmation is shown in Sec.~\ref{sec:data}. The confirmed Cetus stream members and associated stellar debris are discussed in Sec.~\ref{sec:cetus}. The orbital properties of different Cetus components are compared to the current N-body model in Sec.~\ref{sec:model}. We then estimate the mass of the Cetus progenitor dwarf galaxy in Sec.~\ref{sec:mass}. Discussions and conclusion are given in Sec.~\ref{sec:con}.

\section{Algorithmic detection of the Cetus stream}
\label{sec:method}

\subsection{The \texttt{STREAMFINDER} sample}
To search for the Cetus debris over the full sky, we first apply \texttt{STREAMFINDER} to the $Gaia$ EDR3 catalogue. The overall procedure for detecting this stream is similar to the one employed in \cite{ibata21}, however, we made changes in some of the parameters so as to specifically search for Cetus. We use a fixed stellar population template of age = 12.5 Gyr and $\FeH=-2.2$ from the PAdova and Trieste Stellar Evolution Code (PARSEC) library \citep{bressan12}. We adopt a stream width of (Gaussian) dispersion $500\pc$, and allow to search for  neighbors in $10\deg$ along the orbit in a distance range from 10 to $100\kpc$. For comparison, the standard \texttt{STREAMFINDER} run designed to search for thin and cold streams, uses a stream width of $50\pc$, and a distance range of 1--$30\kpc$ \citep{ibata19a, ibata21}. All parameters in this work are motivated by the previous knowledge that we possess for the Cetus system from \citetalias{yuan19}, i.e., Cetus is a fairly wide stream, has distant members all the way to $50\kpc$, and has a low average metallicity ($\FeH\approx-2.0$). The rest of the algorithm is set up to work as described in \citet{ibata21}. It avoids the Galactic disk region ($|b|\leqslant20^{\circ}$) that is prohibitively expensive in computing time to explore, and scans through the heliocentric radial velocity space. Given the measured on-sky position, proper motion, and the assumed distance of a star, its trial orbits are calculated for a grid of radial velocities. The algorithm then evaluates the likelihood of this specific star being in a stream \citep[see Eq (2) in][]{ibata21}. The radial velocity solution with the highest likelihood is selected and is used to calculate the significance that this star belongs to a stream.

In the rest of the paper, we restrict ourselves to stars with a significance $\geqslant6\sigma$ (log-likelihood $\geqslant$ 19.8). Note that this significance cut is lower than that used in previous \texttt{STREAMFINDER} studies: 7$\sigma$ \citep{ibata21} and 10$\sigma$ \citep{malhan21}. The lower significance cut allows us to retain a generous sample (175,514 stars) while the following application of \texttt{StarGO} will help screen spurious members that could have made it into the sample.

\subsection{Application of \texttt{StarGO}}

The next step of the workflow is to apply \texttt{StarGO} to identify DTGs from the orbital properties of the sample stars, as inferred by \texttt{STREAMFINDER}. In particular, we focus on the space defined by the orbital energy, $E$, the orbit's angular momentum along the Galactic $z$ direction, $L_{\rm z}$, and the two parameters $\theta = \arctan(L_{\rm x}/L_{\rm y})$ and $\phi = \arcsin(L_{\rm z}/L$), where $L_{\rm x}$ and $L_{\rm y}$ are the components of the angular momentum vector along the Galactic $x$ and $y$ directions, respectively. Although orbital poles change over time especially in an axisymmetric potential, the changes can remain coherent for a long period of time, and the clustering features of stars from a common origin are mainly preserved. This input space is similar to that used in the previous \texttt{StarGO} applications \citep{yuan19, yuan20b, yuan20a}; we only replace the total angular momentum by $L_{\rm z}$ in the current application. To enhance the signal from the Cetus stream, the sample is further culled by requiring $-45^{\circ}\leqslant\theta\leqslant 0^{\circ}$ from the full range of [$-90^{\circ}$, $+90^{\circ}$], and orbital energy $E<0$. These selections are based on the current knowledge \citep{yuan19,chang20} that the orbit of the Cetus stream is close to polar, and centered on $\theta\approx-30^{\circ}$ (prograde with an angle of $60^{\circ}$ with respect to the Galactic $z$ axis). This yields a sample of 35,286 stars.

\begin{figure*}
\centering
\includegraphics[width=\linewidth]{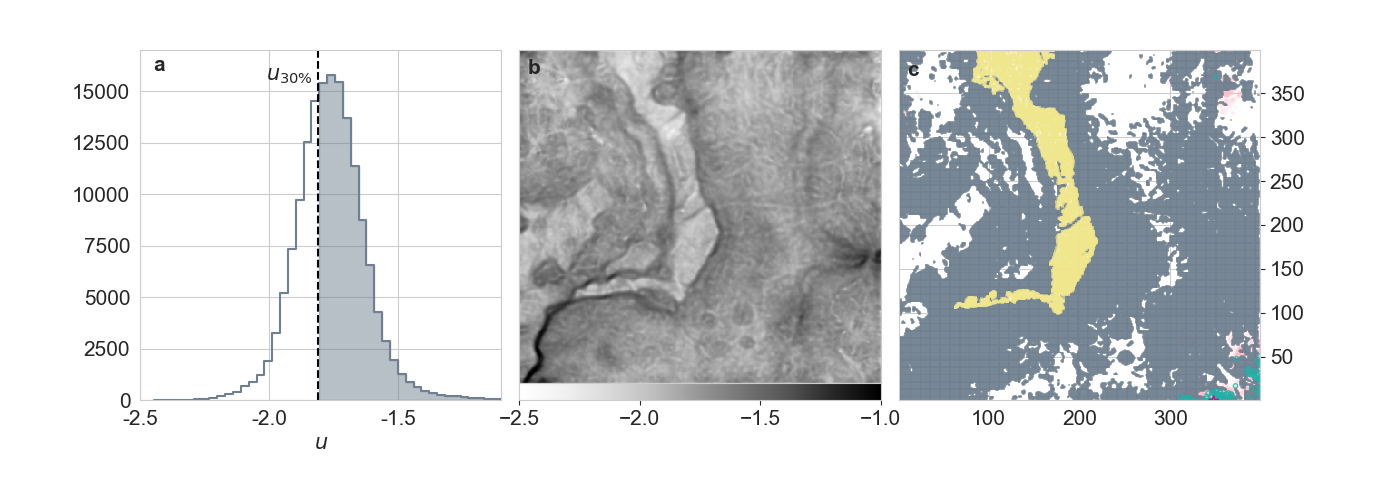}
\caption{Training results from the application of \texttt{StarGO} to the selected \texttt{STREAMFINDER} catalog in the normalized space of ($E$, $L_{\rm z}$, $\theta$, $\phi$). The (4-D distances) differences in weight vectors between neighboring neurons are denoted by the $u$ values, shown as the histogram in (a). The threshold for group identification is $u_{\rm thr}$ = $u_{30\%}$, which defines the division line between the white and shaded areas under the curve. (b): The resulting self-organizing map (400$\times$400 neuron network) color coded by the $u$ values, where the relatively white patches correspond to the stars clustered in the input space. (c) Neighboring neurons with $u\geqslant u_{30\%}$ are colored in grey. At this threshold, two groups are identified in the khaki region. The Cetus members (green circles) identified from \citetalias{yuan19} are mapped to the bottom right corner (see Fig.~\ref{fig:zoom} for a zoom-in view). }
\label{fig:som}
\end{figure*}

The first step of \texttt{StarGO} is to ``feed'' the sample to a 400$\times$400 neural network, whose visualisation is shown in Figure~\ref{fig:som}, panels (b) and (c). At each grid point of the 2-D map, there is a neuron that carries a weight vector with the same dimension as the input vector (i.e. 4-D). The neurons will learn the behavior of the input dataset by iteratively updating their weight vectors, until convergence is reached. The learning algorithm is a self-organizing map (SOM), which preserves the topological structures of the $n$-D input dataset and stores them on the 2-D map \citep{kohonen82}. The difference in weight vectors between neighboring neurons is denoted by a 400$\times$400 matrix, $\mathbf u_{\rm mtx}$. Although the difference is calculated as the distance between neurons in the 4-D weight vector space, the information in each dimension is preserved by their relative locations on SOM. Therefore, clustering algorithm based on SOM does not have the ``curse of dimensionality'' problem that traditional distance-based clustering methods have \citep{lodzis}, with similar distances between different pairs of input data. Compared to the density-based clustering methods, SOM can reveal clusters that have a variety of topologies that differ from a centrally distributed blob, such as a tire tube of linked data point in the n-D input space. 

We are able to get the distribution of all the element $u$ values from $\mathbf u_{\rm mtx}$, shown in Fig.~\ref{fig:som} (a). Neurons with $u\leqslant u_{30\%}$ (30$^{th}$ percentile of the distribution of $u$) have similarities in weight vectors that lie in the top 30\%. These are highlighted in white in (a) and (c), in contrast to the rest of the distribution, shown in grey. These neurons correspond to stars relatively clustered in the input space compared to the rest of the sample. We then create direct linkages between stars and neurons. This is done by finding the neuron that has the weight vector closest to the input vector of a given star, and this neuron is defined as the best matching unit (BMU). Through this step, stars mapped to the islands separated by the grey boundaries in panel (c) are defined as candidate groups at $u_{\rm thr}$ = $u_{30\%}$. At each threshold value $u_{\rm thr}$, different candidate groups can be identified from a SOM. A candidate group is validated according to its estimated contamination rate, which will be discussed in detail in the next section.

\begin{figure*}
\centering
\includegraphics[width=\linewidth]{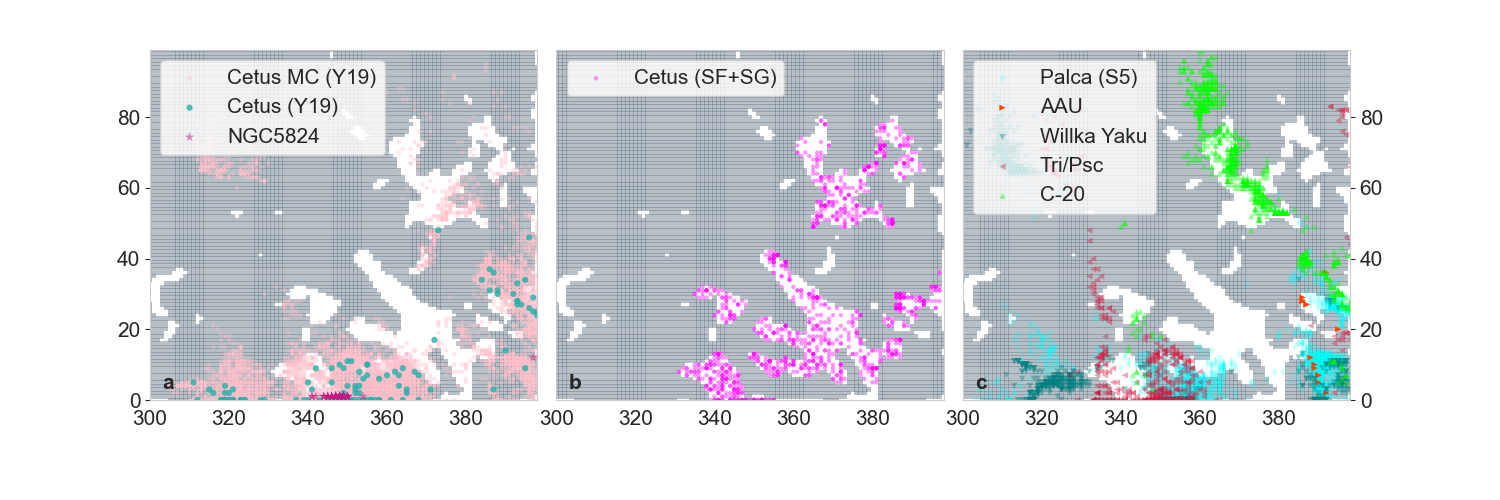}
\caption{Zoom-in view of the SOM in the bottom-right corner of  Figure~\ref{fig:som}, where the previous Cetus members from \citetalias{yuan19} are mapped. Green circles: the 130 Cetus members stars; pink circles: all the Monte Carlo realizations (100 for each) of the Cetus members; violet star: the Monte Carlo realizations of NGC 5824 . (b) The DTGs are identified at $u_{\rm thr}$ = $u_{30\%}$ in the Cetus region, and the stars from the \texttt{STREAMFINDER} sample associated to the Cetus DTGs are plotted by magenta circles. (c) The Monte Carlo realizations of the 24 Palca members (cyan diamonds), 18 AAU members (orange right triangles), 11 Tri/Psc members (red left triangles), 9 Willka Yaku members (dark green down triangles) and 9 C-20 members (light green upper triangles) are mapped to SOM. Except for AAU, most of the realizations of the other stellar debris are located in the Cetus region.}
\label{fig:zoom}
\end{figure*} 

\begin{figure*}
\centering
\includegraphics[width=\linewidth]{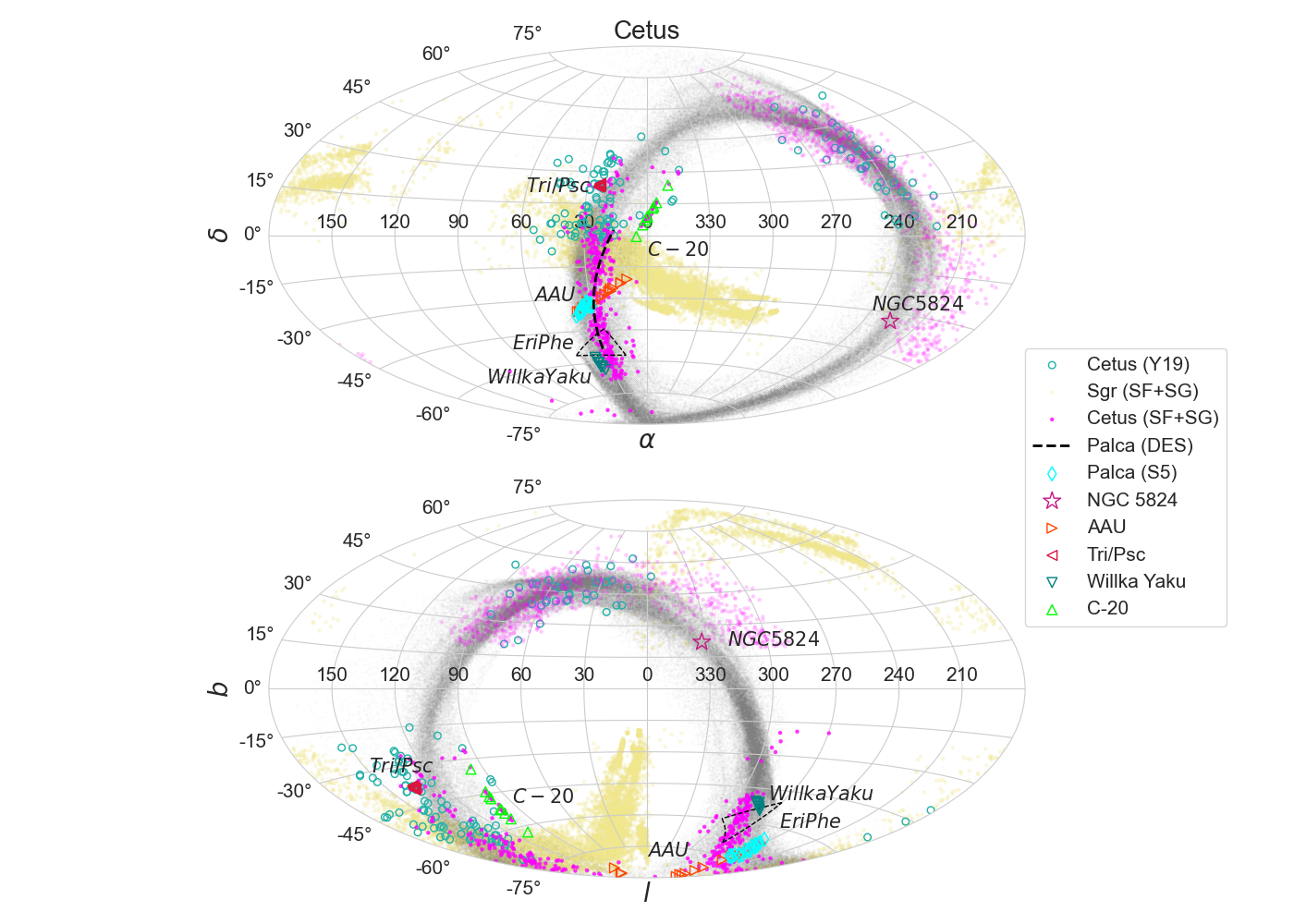}
\caption{On-sky projection of the Cetus system in Equatorial (upper) and Galactic (lower) coordinates. The Cetus stream from \citetalias{yuan19} is denoted by open green circles. The Sgr DTGs identified in this work are shown as khaki circles. The N-body model of the Cetus stream from \citet{chang20} is plotted as gray scatter in the background. The Cetus DTGs identified at the same $u_{\rm thr}$ are denoted as solid magenta circles in the South, and as transparent magenta circles in the North, where it is heavily contaminated. The track of Palca from DES is denoted by the black dashed line. The possible associated debris are represented by a violet star (NGC 5824), cyan diamonds (Palca), red left triangles (Tri/Psc), dark green down triangles (Willka Yaku), and light green upper triangles (C-20). AAU (orange right triangles) is located in the footprint of the southern Cetus, but its association is much weaker compared to the other debris. The region of the EriPhe overdensity is shown as the black dashed triangle. }
\label{fig:onsky}
\end{figure*} 
\begin{table*}[ht]
\begin{center}
\caption{Stellar Stream Systems}\label{tab:dtg}
\bgroup
\def\arraystretch{1.2}
	\begin{tabular}{|p{1.5cm}|p{3cm}p{1.5cm}p{2cm}|p{1.5cm}|p{2cm}p{1.cm}p{2cm}|}
		\hline
\multicolumn{4}{|c|}{Cetus}&\multicolumn{4}{|c|}{Sagittarius}\\ 
  \hline
&& $n$ & Contamination & & & $n$ & Contamination\\
\hline
\multirow{2}{*}{Set} & North& 829 & 36\% & \multirow{2}{*}{Set} &  North&1422 & 35\%\\
& South& 359 & 13\% &&South&7074 &27\% \\
\hline
&& $n$ ($n_{\rm tot}$) & Confidence & &  & & Confidence\\
\hline
\multirow{4}{*}{Part} &Cetus (\citetalias{yuan19}) & 104 (130)& 30\% &  \multirow{8}{*}{Associate} &Arp2 &&  100\%  \\
&Cetus (SF+SG; RV) & 41 (44) & 35\% &&Terzan8 && 100\%\\
&Palca (S5) & 23 (24) & 15\% && Pal12 && 100\% \\
\cline{1-4} 
\multirow{4}{*}{Associate} &NGC5824 & & 86\% &  & M54 && 99\%\\
&C-20 & 9 (9) & 36\%  &&  Whiting1 && 95\%\\
&Tri/Psc & 9 (11)& 19\% && Terzan7 && 42\%\\
&Willka Yaku & 9 (9) & 13\%&&NGC2419 && 10\%\\
\cline{1-4} 
non-Associate&AAU & 2 (17) & 2\%&&Pal2&& 5\%\\
\cline{5-8} 
&&&&non-Associate&Pal4&& 1\%\\
\hline
\end{tabular}
\egroup
\end{center}
Note -- The northern and southern sets of both streams identified in this work are listed in the first part of the table. The Cetus stream identified in different studies are shown in the second part of the table for the Cetus columns. All the candidate associated structures in this study are listed, and divided into the valid ($\mathcal{C}\geq$5\%) and non-valid categories based on their confidence values.
\end{table*}

\subsection{Group identification from known Cetus members}
\label{subsec:gi}

To isolate the parts of the SOM that correspond to likely Cetus members, we modified the group-identification procedure from \citet{yuan20a} by mapping the Cetus stream stars detected by \citetalias{yuan19} to the SOM presented in the previous section. This allows us to use known Cetus members to guide group identification. We see that previously identified Cetus members from \citetalias{yuan19}, shown as green circles in Fig. \ref{fig:som} (c), cluster in the lower right corner. We therefore focus on this region for group identification (see the zoom-in view of the SOM in Fig.~\ref{fig:zoom} a). This is done by decreasing $u_{\rm thr}$ until isolated islands emerge from the gray boundary. We detect a group of islands at $u_{\rm thr}$ = $u_{30\%}$ (magenta patches in Fig.~\ref{fig:zoom} b). At the same threshold, two large DTGs are identified in the middle of the SOM (yellow patches in Fig.~\ref{fig:som} c), which are the most obvious structures revealed in (b). The on-sky projection of these DTGs immediately show that they correspond to the Sgr stream (see yellow scatter in Fig.~\ref{fig:onsky}). On the contrary, the Cetus DTGs (magenta points in panel b) overlap and extend the known Cetus stream (filled green squares in panel a). Both the identified Sgr and Cetus streams have two parts separated by the Galactic plane, since we avoid the disk region ($|b|\leqslant 20^{\circ}$).

As with the previous exploration of the \texttt{STREAMFINDER} catalogue \citep[see Fig. 5 in][]{ibata19a}, we find a broad feature of unknown origin in the region $-60\deg<l<60\deg$ and $-45\deg<b<45\deg$. In the present study, this feature is present well beyond 10 kpc, which was the distance upper limit in the maps of \citet{ibata19a}. This coherent structure surrounds the MW center and reaches as far as 30 kpc, which forms a significant contaminating population for stream identification. Note that the northern Cetus largely overlaps with the footprint of this structure. Therefore, we expect the northern Cetus members to be more contaminated by this halo population, which can be seen from their more diffuse distribution compared to the southern counter part. Due to this reason, we divide all the DTG members into the northern and southern sets, and estimate the contamination fraction ($\mathcal{F}_c$) and significance for these two sets separately. 

To validate the DTGs and assess their contamination levels, we generate a mock sample that has the same distribution as the training sample in the input space, but contains no correlations between the input dimensions from streams. In other words, we reshuffle the training sample $\mathcal{T}$ in each dimension of the input space, yielding a shuffled mock sample $\mathcal{M}$. In doing so, we wash out the correlations that are intrinsically present in the DTGs of the input space and $\mathcal{M}$ can be used to estimate the expected contamination from a smooth halo sample. Compared to the procedure described in \citet{yuan20a}, the two sets of Cetus DTGs (northern and southern) are combined into one group, and similarly for the Sgr DTGs. For a given set $\mathcal{S}$ of $n$ stars in one group identified from the training sample ($\mathcal{T}$) of $N$ stars in the same set, we apply the following steps:

\begin{enumerate}
\item Find the best matching unit (neuron), BMU, for every star of $\mathcal{M}$ on the trained neuron map, and obtain the $n_{\mathcal M}$ stars associated with set $\mathcal{S}$. The probability of stars from $\mathcal{M}$ to be identified in $\mathcal{S}$ is $p_{\mathcal M} = n_{\mathcal M}/ N_{\mathcal M}$.

\item Calculate the binomial probability $\mathcal{P}$ of detecting a set with more than $n$ stars from the total sample of $N$ stars, given probability $p$. If 1 - $\mathcal{P}$ $\geqslant$ 99.73$\%$, the significance of $\mathcal{S}$ is greater than 3$\sigma$, and we consider it as a potentially detected set.

\item If $\mathcal{S}$ is a potentially detected set, we estimate the contamination fraction from $\mathcal{M}$, which is defined as $\mathcal{F}_{c} = p_{\mathcal M}/p$. 
\end{enumerate}

The two sets in the Sgr and Cetus groups all have significance values greater than 3$\sigma$, and their contamination fractions are listed in Table~\ref{tab:dtg}. As expected, the northern Cetus set is much more contaminated ($\mathcal{F}_{c}$ = 36\%) than the southern set ($\mathcal{F}_{c}$ = 13\%), because the coherent halo structure mentioned above heavily overlaps the northern Cetus set, and has a significant contribution to the re-shuffled sample. The two Sgr sets have similar contamination levels, with $\mathcal{F}_{c}$ = 35\% (N) and 27\% (S), and exhibit clear features of the stream from their on-sky projections, shown in Fig.~\ref{fig:onsky}. Based on the comparisons of $\mathcal{F}_{c}$ between the different sets, we are highly confident about the quality of the southern Cetus set, while the northern Cetus set is more prone to biases because of the contamination. We therefore mainly focus on the southern part of the Cetus stream in the rest of the analysis.

\begin{figure*}
\centering
\includegraphics[width=\linewidth]{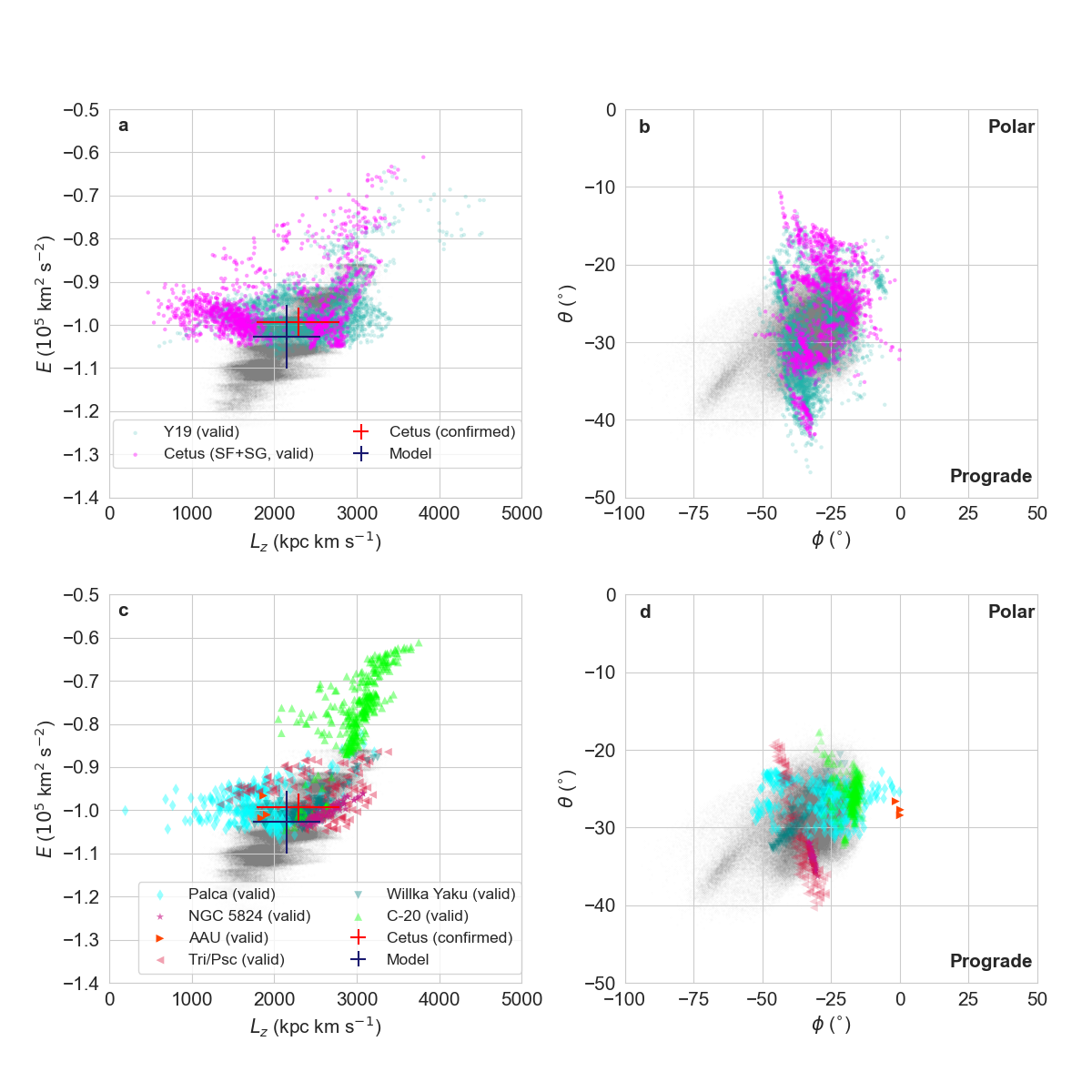}
\caption{Valid MC realizations of the Cetus members with radial velocity measurements that are re-associated with the Cetus DTGs in \ref{fig:zoom} b are plotted in the input space ($L_{\rm z}$, $E$) shown in (a) \& (c), and ($\theta$, $\phi$) shown in (b) \& (d), with the same color coding as Fig.~\ref{fig:zoom}. $\theta$ and $\phi$ represent the two angles of the angular momentum vector, where negative $\theta$ corresponds to a prograde orbit. The existing and new Cetus members overlap with each other in (a) (b), as well as with the Cetus model shown as gray scatter in the background. The red symbol represents the mean and intrinsic dispersion of $L_{\rm z}$ and $E$ from all the confirmed Cetus members, which agree with the values derived from the Cetus model shown as dark blue symbol.
The Cetus members have orbital poles centered around $\theta$ = $-$30$^{\circ}$, \ie 60$^{\circ}$ w.r.t the Galactic-$z$. The valid realizations of NGC 5824, the Palca members, the Tri/Psc members, the Willka Yaku members, and the C-20 members are heavily overlapped in the Cetus region, whereas there are few valid realizations of the AAU members.}
\label{fig:EL}
\end{figure*}

\section{Data}
\label{sec:data}

\subsection{Spectroscopic Data}
\label{subsec:spec}
The list of southern Cetus members was built without any radial velocity information, which gives us the opportunity to check the reliability of these members by gathering radial velocities from public archives and dedicated observations. We are first able to collect velocity measurements for 23 stars by cross-matching with public spectroscopic surveys. We find that 11 stars in our sample were also observed in the SDSS/SEGUE survey \citep{yanny09}, 3 have velocities in LAMOST DR7 \citep{cui12,zhao12}, and 9 stars are in the S5 DR1 \citep{li19}.

To complement this data set we obtained radial velocities from VLT/UVES spectra observed in 18--21 May 2021 and 23--24 October 2021 via programs 0105.B-0235(A) (PI Ibata) and 0108.B-0431(A) (PI Yuan). In total, we obtained spectra for 4 member stars of the C-20 stream \citep{ibata21} in May 2021 (which we'll show below is a cold stream likely associated to Cetus), along with 5 additional members and 21 Cetus stream members in October. The UVES spectrograph was setup with the DIC2 dichroic beamsplitter in the wavelength range of ``437+760", which covers 3730–4990 \AA\ and 5650–9460 \AA. To increase the efficiency of our short exposures of 10--20\,minutes, we used the 2 × 2 pixel binning readout mode and a $1"$ slit, yielding a resolution of R $\sim$ 40,000. All of the spectra were extracted and wavelength-calibrated with the \texttt{esorefex} pipeline\footnote{https://www.eso.org/sci/software/esoreflex/}. To measure the radial velocities of the observed stars, we used the \texttt{fxcor} algorithm in IRAF, cross-correlating against the spectrum of the radial velocity standard star HD\,182572. The metallicities of the three brightest C-20 stars are analysed with \texttt{MyGIsFOS} \citep[see][for more detail]{sbordone14}.

\subsection{Distance Estimation}
\label{subsec:dist}

For all Cetus members detected in this work, we infer their distances using a Bayesian approach following \citet{sestito19, sestito20}. Very briefly, we calculate the probability distribution function (PDF) of the heliocentric distance by merging the Gaia EDR3 photometry ($G$, $BP$, and $RP$) and parallax $\varpi$ with a prior on the Galactic stellar density profile and with \texttt{PARSEC} isochrones \citep{bressan12}. The isochrones are selected to be very metal poor ($\FeH=-2.0$) and with an age of 12 Gyr, in line with the expected properties of an old, low-mass dwarf galaxy like Cetus. Note that the differences between the isochrones used in the \texttt{STREAMFINDER} algorithm and the distance estimation here are negligible compared to the uncertainties of the photometric measurements. This isochrone is used throughout this work. Gaia EDR3 photometry was de-reddened using the dust map from \citet{schlegel98} corrected by \citet{schlafly11}, and updated for the Gaia passbands\footnote{$A_G/A_V = 0.86117$, $A_{BP} /A_V = 1.06126$, and $A_{RP}/A_V = 0.64753$.}. For many objects in our sample, the Gaia parallax values are very uncertain (e.g. $\delta_{\varpi}/\varpi\geq 20$ percent, or $\varpi<0$ mas), which usually implies a large distance and stars that are likely to be associated with a distant halo structure. We therefore assume that all stars are giants, which means that, in cases where the PDF of a star displays both a dwarf and a giant solution, the farthest one is adopted.

There are several distant streams which will be shown to be likely associated to the Cetus system. Since they are in deep photometric surveys, we decide to take the distances estimated from their photometry. For the Palca stream discovered from the DES photometry \citep{shipp18}, we estimate its distances as the average of its six BHB members (33.2 kpc) following the approach presented in \citet{deason11}. Compared to the distance modulus estimation based on the DES photometry \citep[$36\kpc$,][]{shipp18} and the averaged BHB distance estimated from \citep[][d = $36.3\kpc$,]{li21b}, the differences among these estimates are within 10$\%$.

\section{The Cetus Debris System}
\label{sec:cetus}

\subsection{The Original and Expanded Cetus}
\label{subsec:new}

Combining all Cetus DTGs associated with the magenta area in the SOM shown in Fig.~\ref{fig:zoom} (b) there are 359 southern members, 44 of which have radial velocity measurements from spectroscopic surveys and follow-up studies. For each given star, we generate 100 Monte Carlo realizations of dynamical parameters based on observational uncertainties in 6-D space using \texttt{AGAMA} \citep{agama} with the MW potential from \citet{mc17}. We then find the corresponding BMU for each realization on the trained neuron map. The valid realizations are those associated to the Cetus DTGs (magenta patches in Fig.~\ref{fig:zoom} b), shown as magenta circles in the input space (see Fig.~\ref{fig:EL} a, b). 41 out of the 44 candidates that have radial velocity measurements can be associated with the Cetus DTG through with MC procedure and thus are considered as valid members, with an overall confidence of 35\% (35 out of 100 realizations are re-associated). Similarly, the valid realizations drawn from the member list of \citetalias{yuan19} are plotted as green circles and overlap the region of the new members. We then derive the mean and dispersion of all the confirmed Cetus members in $L_{\rm z}$ and $E$ using the formalism of \citet{martin18} by taking into account the observational uncertainties in the radial velocities, proper motions, and distances. The derived values are plotted as the red symbol in Fig.~\ref{fig:EL} ac, which agree with the the values derived directly from the Cetus model from \citet{chang20} shown as the dark blue symbol.

The probability of a given member to be associated with Cetus is the number ratio of the valid associations out of 100 MC realizations, which can be re-associated to the Cetus DTGs on the neuron map. The confidence level ($\mathcal{C}$) is defined as the average probabilities of valid associations for re-associated members. With this definition, 104 of 130 Cetus members listed by \citetalias{yuan19} are identified as members (with non-zero probabilities), with $\mathcal{C}$ = 30\%, denoted as the confidence in Tab.~\ref{tab:dtg}. The fact that the training samples are different in these two studies largely explains why some of the stars previously identified as members by \citetalias{yuan19} are not valid realizations here. Every training set results in a unique SOM, for which the identified groups will not be entirely identical. The results depend on the training set, which in this study is generated from the predicted orbits based on \texttt{STREAMFINDER}. 

We also emphasize that the $\mathcal{C}$ values listed here are indicative as large uncertainties, for instance for the more distant or fainter structures, will lead to low values of $\mathcal{C}$. As such, $\mathcal{C}$ reflects both the accuracy of the orbital parameters and of the association itself and should not simply be taken as a direct evaluation of the intrinsic strength of an association. In this context, we find that all associates of the Sgr stream with $\mathcal{C}\geq5\%$ correspond to known members of the stream \citep{bellazzini20}, using the updated proper motion measurements from \citet{vasiliev21}. We therefore also use this threshold to isolate Cetus associates of interest. Except for the AAU stream, the other associates are considered as valid with $\mathcal{C}$ values listed in Tab.~\ref{tab:dtg}, including NGC 5824 as a highly confident one ($\mathcal{C}$ = 86\%).

\begin{figure*}
\centering
\includegraphics[width=0.95\linewidth]{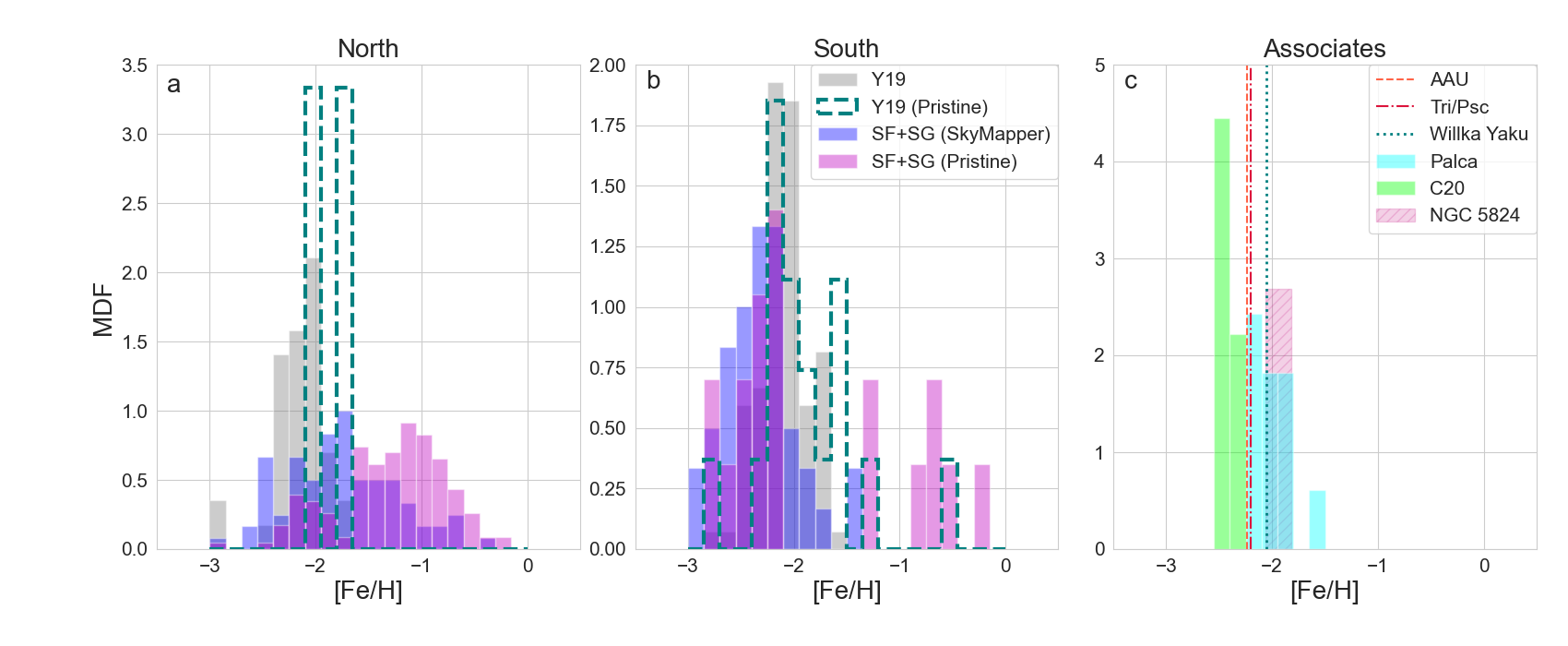}
\caption{MDFs of the Cetus system from different survey data and spectroscopic follow-up studies. (a) The MDFs of the southern Cetus: the spectroscopic metallicities from LAMOST K Giants and SDSS BHBs for \citetalias{yuan19} members (gray); the Pristine metallicities for \citetalias{yuan19} members (dark green dashed line); the metallicities of members identified in this work by cross-matching with Pristine (magenta) and SkyMapper DR2 (blue). The four samples are all consistent, and gives an average $\FeH=-2.1\pm$0.2. The northern Cetus members from \citetalias{yuan19} are very metal poor, whereas the members in this work have a wide MDF and cover the metal rich regime in (b), indicating that they are heavily contaminated in the \texttt{STREAMFINDER} catalog. (c) The metallicities of the stellar debris possibly associated with the Cetus system: NGC 5824 (violet hatched bar and dashed lines); Palca (cyan histogram); C-20 (light green hisogram); Tri/Psc (red dash-dot line), Willka Yaku (dark green dotted line), AAU (orange dashed line) is not strongly associated with the Cetus system.}
\label{fig:met}
\end{figure*}

\subsection{Palca \& Atlas-Aliqa Uma \& Eridanus–Phoenix}
\label{subsec:palca}

We can immediately see from the on-sky projection of the Cetus system shown in Fig.~\ref{fig:onsky} that the new southern extent of Cetus (magenta circles) connects with the previously known Cetus stream (green circles), and overlaps with the Palca stream track (dashed lines) from the DES \citep{shipp18}. The Palca stream has a fairly large width, and extends over 60$^{\circ}$, almost reaching the edge of the DES survey at $\delta = -60^{\circ}$. The potential connection between the Palca and Cetus streams was already discussed by \citet{chang20} and \citet{li21b}. Using the 24 Palca members with radial velocity measurements from S5 (cyan diamonds in Fig.~\ref{fig:onsky}) in the field of AAU \citep{li21a}, we apply the technique described in Sec.~\ref{subsec:new} to quantify the confidence of their association. It is clear that most of the Palca stars are located in the region of the Cetus DTGs in the zoom-in view of the SOM in Fig.~\ref{fig:som} (c). In total, 23 out of 24 Palca members are identified to be associated with the Cetus DTGs, with an average confidence (probability) of 15\% (see Tab.~\ref{tab:dtg}), after assigning an overly generous 20\% uncertainty on this distance estimates we infer for the Palca stars (see Sec.~\ref{subsec:dist}). From the input space, the valid Palca realizations are located well within the region defined by the Cetus members (see Fig.\ref{fig:EL} c \& d). A more detailed comparison between the Palca stream and the extent of the Cetus stream detected in this work is presented in Sec.~\ref{sec:model}.

As shown by \citet{li21b}, the AAU and Palca streams overlap each other in the ($L_{\rm z}$, $E$) space, with a slight difference in the longitudinal angle of orbital poles (\ie $\phi$). This is consistent with this study, where most of the AAU realizations (maroon right triangles) are not mapped in the zoom-in Cetus region in Fig.~\ref{fig:zoom} (c). There are only 3 out of 100 realizations that are associated with the Cetus DTGs, shown in Fig.\ref{fig:EL}. Their association has confidence $\mathcal{C}$ = 2\%, and considered as invalid based on the threshold of valid association defined in this study (see details in Sec.~\ref{subsec:new}).

Among the southern stellar structures discovered in the DES, there is an overdensity, Eridanus-Phoenix (EriPhe) centered at $l$ $\approx$ 285$^{\circ}$, $b$ $\approx$ $-$60$^{\circ}$ \citep{li16}. We show it is right in the footprint of the southern Cetus, denoted by the dashed black triangle in Fig.~\ref{fig:onsky}. There are no spectroscopic follow-up observations of the EriPhe members. However, \citet{chang20} predicted its possible association with the Cetus stream because the simulated stream covers the distance range of EriPhe ($d\approx 16\kpc$) in the same sky area. In this work, the southern Cetus members detected in the region of EriPhe have similar distances, which further strengthen this association (see details in Sec.~\ref{subsec:cetus-new}). Note that this is one possible scenario for the origin of EriPhe, although other scenarios are also discussed in \citet{li16, donlon20}, which is before the discovery of the southern Cetus stream.

\subsection{Triangulum/Pisces \& Willka Yaku}

Besides the kinematically hot Palca stream, there are three cold streams \citep[Tri/Psc;][]{bonaca12, martin13}, Turbio, and Willka Yaku \citep{shipp18} that are suggested to be associated with the Cetus stream by \citet{bonaca21}. Among them, Tri/Psc and Willka Yaku are shown to be very close to NGC 5824 in phase space by \citet{li21b}, who claims that they likely came from the same group infall. Here, we use the \citet{martin13} member list of Pisces stars from SDSS DR8 to quantify the confidence of this association. We take the distance estimated from the stream's MSTO as seen in the PAndAS survey \citep[$d \sim 27\kpc$][]{martin14}, and assume an uncertainty of 20\%. The analysis yields $\mathcal{C}$ = 19\% for an association to the Cetus system, indicating a strong association.

For the Willka Yaku stream, we take the distance estimate ($d = 36.3\kpc$), and radial velocity measurements of its 9 members from \citet{li21b}. The resulting confidence $\mathcal{C}$ = 13\% is similar to that of the Palca stream. For this stream as well, we reach the conclusion that it is confidently associated to the Cetus system.

\subsection{C-20}
\label{subsec:C-20}

In addition to possible associations already listed in previous studies, we notice a thin stream-like track from the southern Cetus members, at $\alpha\sim0^{\circ}$, $0\deg\lesssim$ $\delta\lesssim 30\deg$ in Fig.~\ref{fig:onsky}. This stream track coincides with the C-20 stream discovered by \citet{ibata21} and is relatively thin and cold compared to the Cetus stream. There are 14 C-20 stars in common with the Cetus member list. We obtained accurate velocity measurements for nine of those C-20 stars, which allows us to determine an association between the two structures at relatively high confidence ($\mathcal{C}$ = 36\%). This is visible in the mappings of the MC realizations of these nine C-20 stars (light green triangles in Fig.~\ref{fig:som} c), most of which are located in one of the Cetus DTGs. Consistently, the valid realizations of the C-20 stars heavily overlap with the Cetus members in the input space (see Fig.~\ref{fig:EL} c \& d).

\subsection{Metallicities of the different components}
\label{subsec:met}

We are able to compare the metallicities of all Cetus components and possible associations by cross-matching with data from different surveys. Fig.~\ref{fig:met} (a) \& (b) show the metallicity distribution function (MDF) of the Cetus members from \citetalias{yuan19} and of the new members in this study, divided between the northern and southern sets. The MDF in \citetalias{yuan19} from LAMOST K Giants and SDSS BHBs represented by the gray histograms and give an average [Fe/H] = $-$2.2 (North) and $-$2.3 (South). For those southern members also present within the footprint of the Pristine survey \citep{starkenburg17}, we rely on the photometric metallicities of this survey, based on narrow-band $CaHK$ photometry. From the member list detected in this study, there are 154 northern stars and 20 southern ones in the Pristine survey, the MDFs of which are shown as magenta histograms in (a) and (b). The southern set has an average [Fe/H] = $-$1.9, and 13 members are very metal-poor (VMP, [Fe/H] $\leqslant$ $-$2). On the other hand, the northern MDF has a bimodal distribution, which clearly shows a large source of metal rich stars that are likely contaminants (see the discussion in Sec.~\ref{subsec:gi}). Similarly, using the photometric metallicities derived from the SkyMapper DR2 data \citep{huang21a, huang21b}, the northern set shows an extended distribution, and the southern MDF exhibits a tight distribution with an average [Fe/H] = $-$2.3 shown as blue histograms. There are 35 out of 41 stars are VMP in the southern set. In summary, the average metallicity of the southern Cetus is [Fe/H] = $-$2.1$\pm$0.2 from the two photometric data sets above\footnote{Before doing this average, we checked that the offset between Pristine and SkyMapper DR2 metallicities is small (0.07 dex) in the metal poor regime ([Fe/H] $\leqslant$ $-$1.5) from the cross-matched sample.}. Note that the 21 Cetus stars observed by VLT/UVES are chosen from the southern members with photometric metallicities below $-2$. The detailed spectroscopic analysis of element abundances will be presented in the future work\footnote{A preliminary analysis of the iron abundance of two of the stars, Cetus23 ($\FeH=-2.12\pm0.06$) and Cetus24 ($\FeH=-1.96\pm0.09$) confirms their good agreement with the photometric metallicities ($\FeH=-2.4\pm0.2$ and $-1.8\pm0.2$, respectively) and the mean metallicity calculated above.}.

All the new structures we associate with Cetus based on the exploration of the SOM have average metallicities that are in agreement with these values. We show the metallicity of all associated debris in Fig.~\ref{fig:met} (c). The MDF of the 23 Palca members (cyan histogram) has an average $\FeH=-2.02\pm0.04$ from \citet{li21b}. The average metallicities of NGC 5824 is  $\FeH = -1.94\pm0.12$ (violet hatched bar) from \citet{roederer16}, similar to $-$2.11$\pm$0.01 from \citet{mucciarelli18}. The average metallicity of Tri/Psc is $\FeH = -2.2\pm0.3$ from SDSS DR8 spectroscopic data. Willka Yaku has an average metallicity $\FeH= -2.05\pm0.07$ from S5 \citep{li21b}. The metallicities of three C-20 stars that have been derived from the VLT/UVES spectra are denoted by the light green histogram, with an average $\FeH= -2.44$. These two strong associates are both very metal poor, consistent with the mean metallicity of the southern Cetus members. Although the association between the Cetus and the AAU streams is much less obvious, we note that the metallicity of the latter is also compatible, with an average $\FeH=-2.24\pm0.02$ according to \citet{li21a}.

\begin{figure*}
\centering
\includegraphics[width=\linewidth]{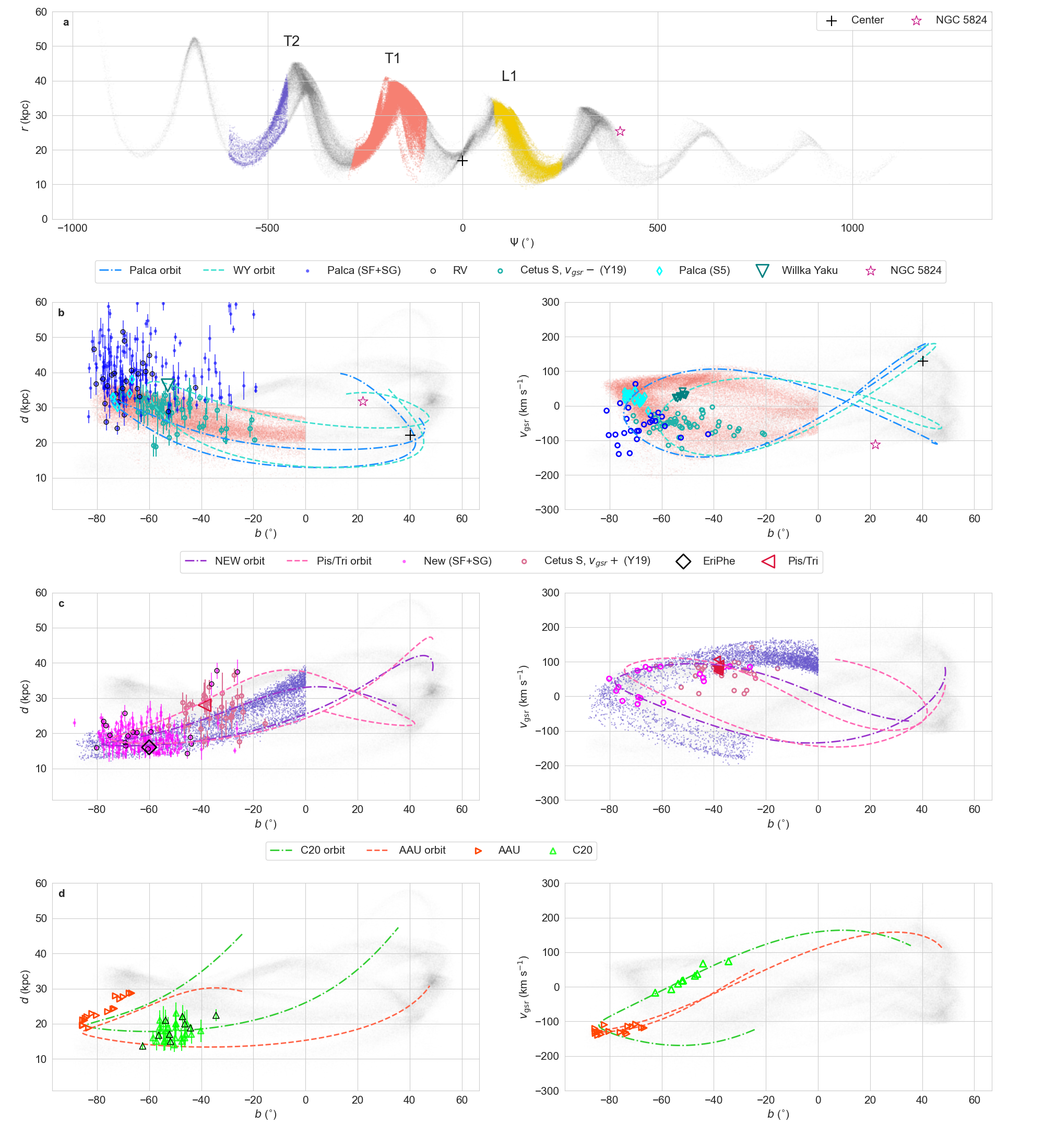}
\vspace*{-12mm}
\caption{Comparisons of the Cetus debris with the stream model from \citet{chang20} in the ($b$, $d$, $v_{\rm gsr}$) space. (a) The entire simulated stream (gray scatter) is unfolded along the azimuthal angle $\Psi$ and shown in the space of ($\Psi$, $r$). The three wraps in the leading and trailing arms closest to the center of the disrupted progenitor are L1 (yellow), T1 (orange), T2 (light purple), and only the segments of the wraps entering the sky region of the observed streams are colored. (b) The Cetus-Palca wrap is made of the northern Cetus with negative $v_{\rm gsr}$ (green circles), the Palca wrap (blue points) including the Palca members in S5 (cyan diamonds). The entire structure follows T1 (orange), as well as its sampling orbit (light blue dash-dot line). The best fit orbit (turquoise dashed line) of the Willka Yaku stream (dark green downwards triangles) are in line with the Cetus-Palca wrap. (c) The Cetus-New wrap is composed of the northern Cetus with positive $v_{\rm gsr}$ (pink circles), the new southern wrap (magenta dots), which follows T2 (light purple), as well as its sampling orbit (purple dash-dot line). The Tri/Psc stream resides in the northern members, with its best fit orbit (pink dashed line) aligned with the new wrap. The EriPhe overdensity (black diamond) has the same distance as the new southern members at $b=-60^{\circ}$. (d) The C-20 stream (best fit orbit: green dash-dot line; members: light green upper triangles) and the AAU stream (best fit orbit: orange; members: orange right triangles) mainly agree with L1 (yellow), despite some offsets in distances.}
\label{fig:wrap}
\end{figure*}

\section{Comparison with the Cetus Model}
\label{sec:model}

We now compare the orbital properties of different Cetus components with simulations. Based on the previously detected Cetus stream in the northern sky, \citet{chang20} explored a range of initial conditions for the progenitor and found a favorable model that can match the morphology and features of the entire stream as seen in the ($b$, $d$, $v_{\rm gsr}$) space. This model is represented by the small light gray points in Fig.~\ref{fig:wrap}. In brief, the system has undergone a very long period of tidal stripping (8 orbital periods, $\sim$ 5 Gyr), and left multiple wraps in the form of both trailing and leading arms. In order to compare the new findings with the N-body model, we unfold the simulated stream in the space of orbital phase and distance, ($\Psi$, $r$), as done by \citet{chang20} and shown in Fig.~\ref{fig:wrap} (a). Here $\Psi$ is the angle between the star and the progenitor’s center with respect to the MW center, and $r$ denotes the Galactocentric distance. The center of the disrupted Cetus progenitor from the model is currently located at (0$^{\circ}$, 20 kpc). We highlight the part of the streams in the Galactic South with different colors in (a), and name the wrap closest to the center as L1 (yellow) in the leading arm ($\Psi\geqslant$ 0$^{\circ}$), T1 (orange) and T2 (light purple) in the trailing arm ($\Psi\leqslant$ 0$^{\circ}$), respectively. 

The Cetus stream previously identified by \citetalias{yuan19} is plotted in (b) \& (c). The southern members are separated into two clumps in ($b$, $v_{\rm gsr}$), with opposite signs in $v_{\rm gsr}$ (negative velocities represented by green circles in b, and positive velocities as pink circles in c). These two clumps are the most densely populated structures in the previous findings and clearly have different gradients in both $d$ and $v_{\rm gsr}$ as a function of $b$. In order to reproduce these features in the model, the center of the disrupted system (black cross) does not overlap the associated globular cluster NGC 5824 (purple star) \citep[see detailed discussions in][]{chang20}. We see that the globular cluster is located in the wrap stripped earlier than L1 in the leading arm. Although the predicted center might shift based on the new findings, the relative location of these wraps along the orbit remain the same. 

\subsection{Cetus-Palca Wrap}
\label{subsec:cetus-palca}
In panels (b), we plot the southern extent of the Cetus stream beyond 30 kpc (blue dots) that are detected in this work. In this relatively distant group with an average distance of 40\kpc, there are 26 stars that have radial velocity measurements. These are represented by the black and blue circles in the two panels of (b). The Palca members \citep[cyan diamonds][]{li21a} have similar distances and radial velocities compared to the members at $b=-70^{\circ}$. To show the orbit of this distant group, we adopt the orbit-sampling procedure instead of orbit-fitting because the latter would have been a poor approximation for such streams that are dynamically hot and physically broad. We use the phase-space information of the 26 stars with velocity measurements to constrain its orbit. The orbit of each member is obtained from 200 samplings of the observational uncertainties in proper motion, radial velocity, and distance. The averaged orbit from the samplings of the 26 members is denoted by the light blue dash-dot line. It has similar gradient as the previous Cetus component with negative $v_{\rm gsr}$ and aligns very well with the T1 wrap (orange). We therefore conclude that this distant group is the Palca stream discovered in the DES \citep{shipp18} and is the southern extent of the Cetus stream with negative $v_{\rm gsr}$, previously identified in the northern sky by \citetalias{yuan19}. The best fit orbit (turquoise dashed line) of the Willka Yaku stream located at $b$ = $-$53$^{\circ}$ (red left triangle) is inline with the Palca orbit. We name this entire stream structure in the Galactic South as the Cetus-Palca stream wrap, and its part in the southern sky as the Palca wrap.

\subsection{Cetus-New Wrap}
\label{subsec:cetus-new}

In Fig.~\ref{fig:wrap} (c), the southern Cetus members (magenta dots) at smaller distance ($d \approx 18\kpc$) form a clear stream track in the ($b$, $d$) space. Although this closer group is located in the same region as the Palca wrap on the sky, it is clearly a new wrap because it spreads over a different distance range and has a distinct gradient in ($b$, $d$). Its averaged orbit from sampling 18 members with velocity measurements (magenta circles) is shown as the purple dash-dot line. We find that the new wrap is the southern extension of the previously detected Cetus with positive $v_{\rm gsr}$ and closely follows the T2 wrap from the model (light purple). The EriPhe overdensity (black diamond) sits well within the southern wrap with similar distance. The Tri/Psc stream (red left triangles) resides among the Cetus members at $b\approx-38^{\circ}$, and its best fit orbit (pink dashed line) aligns closely with the orbit of the new wrap. We refer to the whole stream structure as the Cetus-New stream wrap, and its part in the southern sky as the new southern wrap. The Cetus-New wrap is the closest and densest wrap, thus we estimate the properties of the Cetus stream according to this wrap. By applying the formalism of \citet{martin18}, the stream width is estimated to be 5$\deg\pm0.3\deg$ equivalent to the maximum width of $\sim$1.6$\,$kpc at the average distance of 18 kpc. The velocity dispersion is 12$\pm3\,\kms$ based on the 15 members with radial velocity measurements.

\begin{figure*}
\centering
\includegraphics[width=\linewidth]{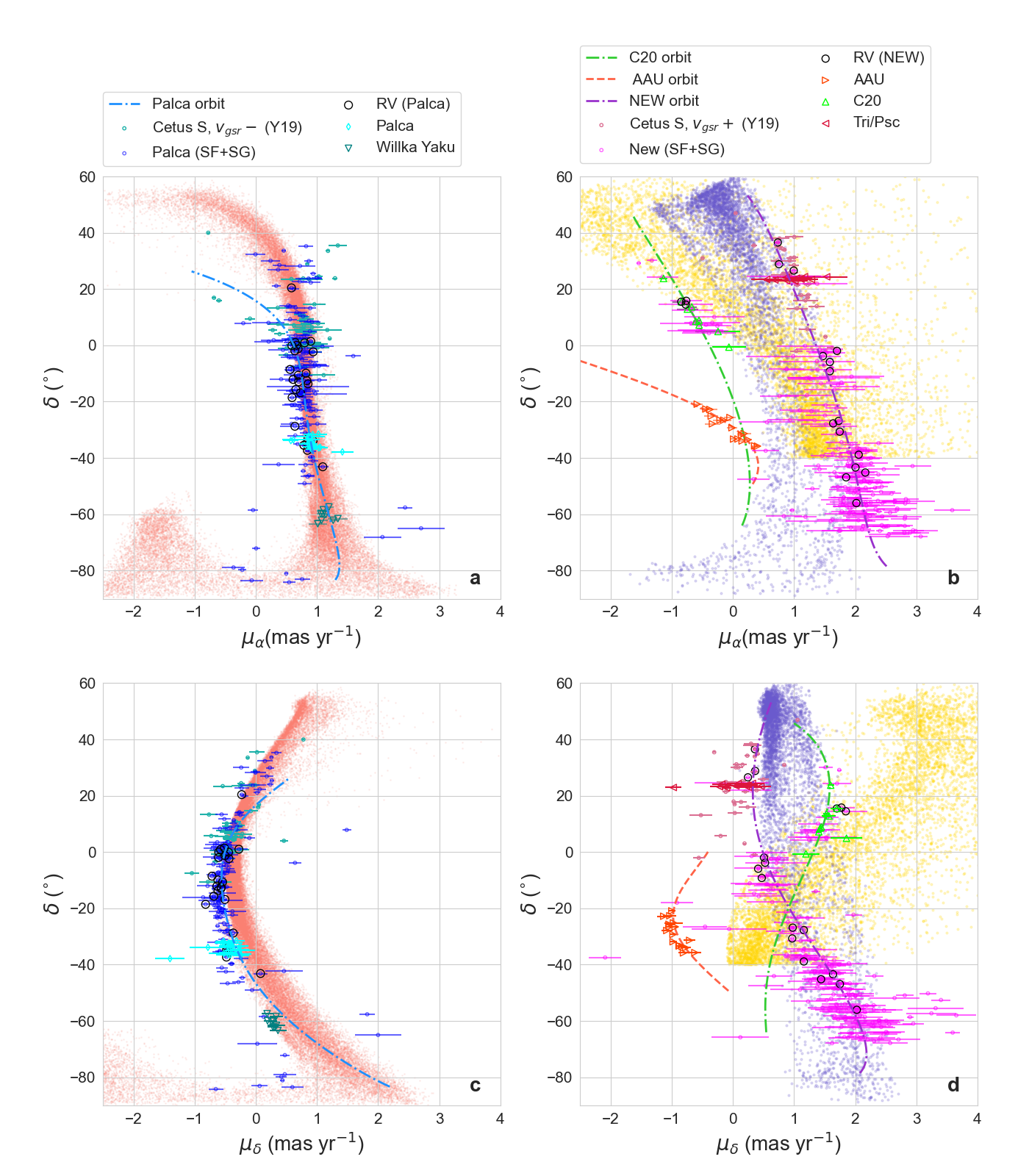}
\vspace*{-6mm}
\caption{Same as Fig.~\ref{fig:wrap} in proper motion spaces, ($\delta$, $\mu_{\alpha}$) and ($\delta$, $\mu_{\delta}$). (a) \& (c): Cetus from \citetalias{yuan19} (negative $v_{\rm gsr}$, green), the Palca wrap detected in this work (blue), and the Palca (cyan diamonds) and Willka Yaku (dark green down triangles) streams from S5 \citep{li21a} form a continuous stream track, and agree with the averaged sampling orbit of the Palca wrap as well as T1 (orange). (b) \& (d): Cetus (positive $v_{\rm gsr}$, pink), the new southern wrap (magenta), and its orbit (purple), and the Tri/Psc stream (red left triangles) align with T2 (light purple). C-20 (light green) and its best fitting orbit (green) generally agree with L1 (yellow), whereas AAU and its fitted orbit (orange) deviate further from L1.}
\label{fig:pm}
\end{figure*} 

\subsection{C-20 Wrap}
\label{subsec:cetus-c20}

The best fit orbit of C-20 is represented by the green dash-dot line in Fig.~\ref{fig:wrap} (d). It is clear that its orbit has a different track from that of Cetus-Palca (T1) and the Cetus-New wrap (T2). The closest wrap of the Cetus model to C-20 is L1 (yellow) in the region with $b<0^{\circ}$ and $\delta>-40^{\circ}$. L1 follows the stream track of C-20 in ($b$, $v_{\rm gsr}$) but has a distance offset of about $10\kpc$ in distance. Interestingly, the orbit of C-20 comes across the AAU stream in both spaces. The best fit orbit of AAU is shown as the orange dashed lines and they also agree with L1 in the observed region of the sky.

\subsection{Proper Motion}
\label{subsec:pm}
We further investigate the correspondence between the different Cetus components and the simulated stream wraps from the Cetus model in proper motion space. In Fig.~\ref{fig:pm} (a) \& (c), the Cetus-Palca wrap spreads from $\delta=40^{\circ}$ to $\delta=-60^{\circ}$, and is made of the previously detected Cetus stream with negative $v_{\rm gsr}$, the Palca wrap defined in this work, the Palca and the Willka Yaku streams from S5 \citep{li21a}. The first wrap in the trailing arm (T1; orange) aligns very well with the entire stream track, as well as with the averaged sampling orbit of the Palca wrap (light blue dash-dot lines). The deviation from the sampling orbit occurs at $\delta>15^{\circ}$, which is likely due to the lack of members with radial velocity information in the region.

The Cetus-New wrap also stretches over 100$^{\circ}$ in the same part of the sky as the Cetus-Palca wrap. The whole wrap aligns with the averaged sampling orbit (purple dash-dot line) of the new southern wrap, and also agrees with the second wrap in the trailing arm (T2; light purple). The Tri/Psc stream is located in the footprint of Cetus in the northern sky and has very similar proper motion measurements to the northern Cetus with positive $v_{\rm gsr}$.

The C-20 stream and its best fit orbit (turquoise line) follow similar tracks to the first wrap in the leading arm (L1; yellow). Although the AAU stream and its best fit orbit (orange line) follows L1 closely in ($b$, $d$, $v_{\rm gsr}$) space, they show different tracks in the proper motion space. We reach the conclusion that we cannot rule out the possibility that the AAU stream is associated to the Cetus system, but this association is weaker compared to that of C-20 with the Cetus stream.

\section{Mass of the Cetus Progenitor}
\label{sec:mass}
\begin{figure*}
\centering
\includegraphics[width=0.95\linewidth]{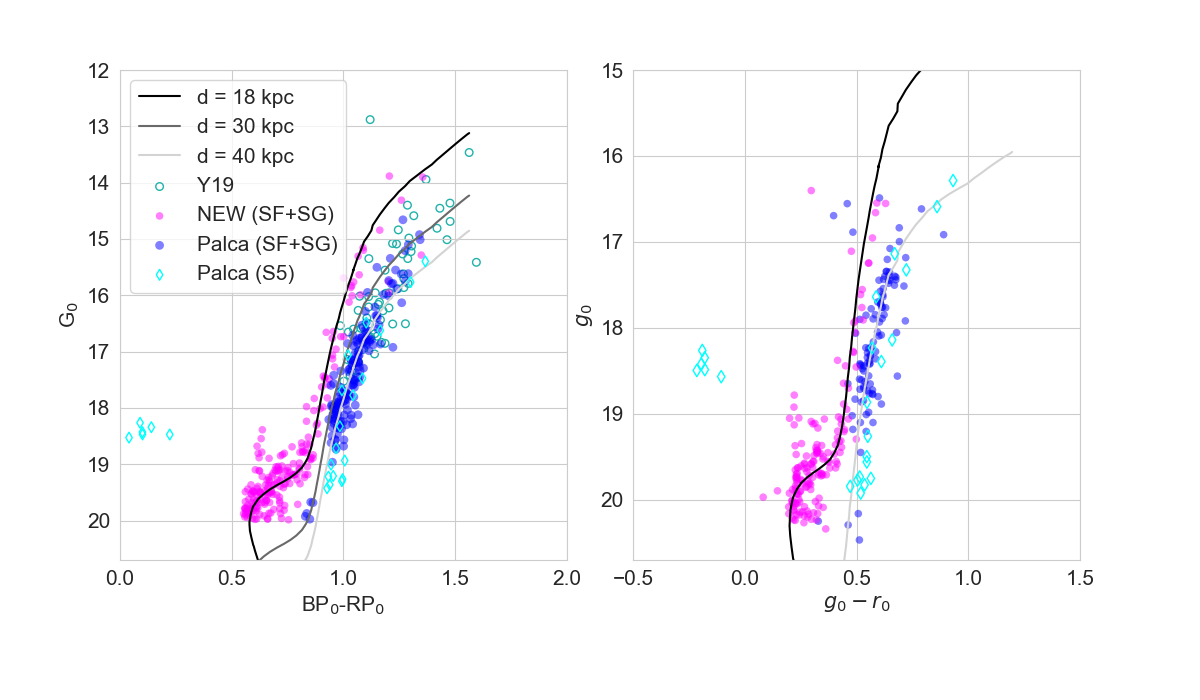}
\caption{CMD of the Cetus members from $Gaia$EDR3 (left) and DES DR2 (right), where the magnitudes and colors are extinction-corrected values. The Palca wrap (blue solid circles) matches the PASEC isochrone (age = 12.5 Gyr, [Fe/H] = $-$2) at d = $40$ kpc, which is the average heliocentric distance of the Palca members estimated. The new southern wrap (magenta solid circles) is consistent with the same isochrone at d = 18 kpc. The previous Cetus K Giant members (open green circles) follows the isochrone at d = 30 kpc.}
\label{fig:cmd}
\end{figure*}

We have shown that the southern part of the Cetus stream detected in this work has two wraps located at different distances. This can also be seen from their color-magnitude diagrams (CMD) in Fig.~\ref{fig:cmd}, where we plot the southern Cetus members using $Gaia$ EDR3 \citep{riello21} and DES DR2 photometry \citep{desdr2}. After correcting for the extinction (see details in Sec.~\ref{subsec:dist}), the Palca (blue) and the new southern wrap (magenta) are consistent with the PARSEC isochrones ([Fe/H] = $-$2, age = 12.5 Gyr) assuming the IMF from \citet{kroupa21, kroupa22}xw at $d = 40\kpc$ (black) and $d = 18\kpc$ (light gray) respectively. These distances are the averages of all members in each of these two wraps estimated from $Gaia$ EDR3 photometry (see details from Sec.~\ref{subsec:dist}). The Palca members from \citet{li21a} are represented by open cyan diamonds and are consistent with the isochrone at $d = 40\kpc$. The previous Cetus members from \citetalias{yuan19} (open green circles) are in agreement with the isochrone (medium gray) at their average distance of $d = 30\kpc$.

\begin{figure*}
\centering
\includegraphics[width=0.95\linewidth]{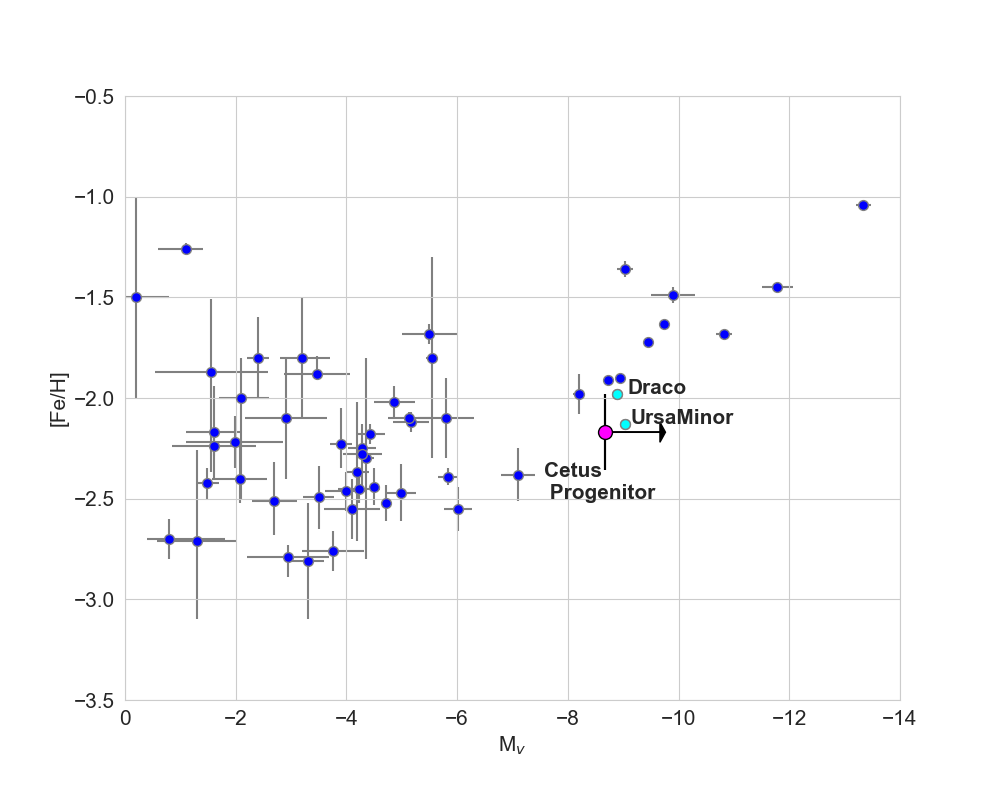}
\caption{Luminosity-metallicity plot of classical dSphs and ultra-faint dwarfs \citep[see][and references within]{battaglia21}. The lower mass limit of the Cetus progenitor is shown as magenta circle with right arrow, which lies in the low-mass classical dSph regime, close to Ursa Minor and Draco dSphs.}
\label{fig:mass}
\end{figure*}

To estimate the minimum total luminosity of the Cetus progenitor, we sum the fluxes of all stars brighter than $G_0 = 20$ in the Palca wrap, and those brighter than $G_0 = 18.5$ in the new southern wrap. For the K giant members from \citetalias{yuan19}, we sum up the flux of those brighter than G$_0$ = 17 mag. We then obtain the correction factor by summing, for each sample, the fluxes of stars fainter than these magnitudes according to the luminosity function of the stellar population that corresponds to the isochrone shown in Fig.~\ref{fig:cmd}. The total corrected luminosity of these two wraps combined with the previous K giant members is $10^{5.4}\lsun$. This gives us a lower mass limit for the Cetus progenitor,  $M_V = -8.7$, and $M_{\ast}$ = $10^{5.6}\msun$, assuming a stellar mass-to-light ratio of 1.6 for dwarf galaxies as given in \citet{woo08}. We then compare this value with the luminosity-metallicity relation of the MW satellite dwarf galaxies in Fig~\ref{fig:mass} \citep[see \eg][and references within]{battaglia21}, where the lower limit on the total magnitude of the Cetus progenitor is denoted by the magenta circle and a right arrow. This lower limit is compatible with the distribution of MW dwarf galaxies in this plane and could imply that the progenitor of Cetus was not significantly more massive than a dwarf spheroidal galaxy (dSph) like Draco or Ursa Minor \citep[$\sim10^{5.7}\msun$][]{kirby13}.

\section{Discussions and Conclusions}
\label{sec:con}

In this study, we made a fusion of \texttt{StreamFinder} and \texttt{StarGO} to search for members of the Cetus system in the all-sky $Gaia$ EDR3 data. We confirm that the Palca stream discovered by \citet{shipp18} and further studied by \citet{chang20} and \citet{li21a} is the southern extension of the Cetus component with negative $v_{\rm gsr}$ detected by \citetalias{yuan19}. We identify 160 candidate members that belong to the Cetus-Palca stream wrap in the distance range of 30 kpc to 60 kpc and in the sky coverage of $-40\deg<\delta<40\deg$. The Willka Yaku stream is shown to be confidently associated with the Cetus-Palca wrap, and extends the structure 20$^{\circ}$ in the southern sky. We also present accurate line-of-sight velocities for 26 stars in this wrap from different spectroscopic surveys and follow-up observations, and show that their orbits are consistent with the first wrap in the trailing arm of the Cetus model of \citet{chang20}.

Furthermore, we identify a second, densely populated southern wrap with 205 stars, overlapping the Palca wrap on the sky, but located much closer, at an average distance of 18\kpc. Based on 18 stars with line-of-sight velocities, we show that these are the extension of the previously detected Cetus stream with positive $v_{\rm gsr}$, and have a strong association with the Tri/Psc stream. The Cetus-New wrap spreads over 100$^{\circ}$ on the sky, and matches perfectly the second wrap in the trailing arm of the Cetus model. 

Our exploration also highlights a thin stream that belongs to the same system in the northern sky. It coincides with the thin stream C-20 discovered by \texttt{STREAMFINDER} in \citet{ibata21}. We confirm that C-20 is dynamically associated with the Cetus system from its nine members with line-of-sight velocity information. Thus, it is the second most confidently associated structure after globular cluster NGC 5824. The best fit orbit of C-20 shows that it was possibly stripped with the first wrap in the leading arm. 

The association between the Cetus system and the ATLAS-Aliqa-Uma stream is weaker compared to the other associations described above. However, the best fit orbit still closely follows the first leading wrap. The observed ``kink'' features and gaps of the AAU stream suggest that it might have been perturbed, potentially by a small mass dark matter halo. \citet{li21a} suggested that it could be due to an encounter with the Sagittarius dwarf galaxy. Given the proximity of the orbits of the AAU and the Cetus system, we propose a scenario where the perturber is the shredded dark matter halo of the Cetus progenitor. We will investigate this possible connection in future studies.

Based on the members of the southern Cetus stream from \citetalias{yuan19} and this study, we measure its average metallicity, $\FeH=-2.17\pm0.2$ and we estimate a lower limit to its total luminosity ($M_V = -8.7$). As such, the Cetus progenitor is compatible with other MW satellite galaxies similar to the Ursa Minor dwarf galaxy with stellar mass $\sim10^{5.7}\msun$. In this case, NGC 5824, as the most confident associate, has a similar stellar mass (10$^6\msun$) to the Cetus progenitor. How such a massive GC is associated with the progenitor system remains puzzle if our estimates are accurate.

The scenario of NGC 5824 being the nuclear star cluster of the Cetus progenitor is disfavored by the N-bdoy modeling of the Cetus stream detected in the northern sky \citep{chang20}. A key conclusion of this work is that the center of the disrupted progenitor cannot be at the location of NGC 5824 in order to populate streams in the detected region instead of the region around NGC 5824. In this work, we identify the two southern wraps as the extent of the Cetus stream that are predicted by the favored model from \citet{chang20}. At the same time, we still do not detect any densely populated structure around NGC 5824, even though the cluster falls within the coverage of the data we used. All of these lines of evidences support the scenario that NGC 5824 is not the core of the Cetus progenitor. On the other hand, the progenitor would need to have been more massive than $10^{6}\msun$ in stellar mass to host NGC 5824 as its nuclear star cluster. According to the nuclei to stellar mass relation given by \citet{georgiev16}, a nuclear star cluster of $\sim10^{6}\msun$ is typically hosted by a galaxy of at least $10^{8}\msun$ and thus much more massive than our estimate of the Cetus progenitor. We therefore conclude that it is unlikely that NGC5824 was the former nucleus or a globular cluster of the Cetus progenitor. However, due to its very similar dynamical properties we think it is probable that it was accreted with the group of satellites that included the Cetus progenitor.

Finally, we also identify another three associated substructures (Tri/Psc, Willka Yaku, C-20) that, given their morphology, are very likely globular cluster streams. They belong to the stream wraps of the Cetus system that are closer to the predicted center compared to NGC 5824 according to the Cetus model. Contrary to what we discussed above for NGC 5824, it appears natural to associate these apparently smaller systems to the progenitor of Cetus. It is likely that their progenitors were less massive than $10^5\msun$, because such GCs typically dissolve within a Hubble time due to internal dynamical effects \citep[see \eg][]{baumgardt03, kruijssen19}. Based on the stellar mass and the N-body model, we estimate a total dark matter halo mass of the Cetus progenitor is around 10$^9\msun$ \citep[see \eg][]{read17}. It is known from observations of galaxies in the Local Universe that the stellar mass of the whole GC system of a galaxy is a factor of $10^{-4}$ smaller than the total halo mass of its galaxy \citep{harris13}. This scenario, in which the progenitor globular clusters have a combined mass of $\sim10^{5}\msun$, is, while speculative, consistent with having been members of the Cetus progenitor.

The Cetus system is a perfect example of a dwarf galaxy that has undergone several orbital periods of stripping and that left behind complex stellar debris in the form of multiple wraps around the Galaxy. It further confirms the complexity of mapping the various structures of the MW stellar halo and how streams that appear separated in the sky, identified from various datasets with different techniques, can actually be produced by the same accretion event. Revealing this common origin strongly benefits from exploring the dynamical properties of their component stars, as we did here by combining \texttt{STREAMFINDER} and \texttt{StarGO}. The N-body simulation tailored to the originally discovered parts of the stream is also a very powerful tool that provides insight into how the different pieces of the debris fit together. Only through such multi-approach studies can we hope to understand how progenitors were shredded during their long disrupting history.

\section*{Data Availability}
We publish the Cetus model from \citet{chang20} used in this study. The catalog of 2$\times10^5$ star particles in the last snapshot of N-body simulation is hosted at \url{https://doi.org/10.5281/zenodo.5771585}.

\section*{Acknowledgements}

Z.Y. wishes to thank Guillaume F. Thomas for helpful comments when drafting the paper. Z.Y., RAI, NFM, AA, BF acknowledge the European Research Council (ERC) under the European Unions Horizon 2020 research and innovation programme (grant agreement No. 834148). Z.Y., RAI, NFM, EC, and PB also acknowledge funding from the Agence Nationale de la Recherche (ANR project ANR-18-CE31-0017). K.M. acknowledges support from the Alexander von Humboldt Foundation at Max-Planck-Institut f\"ur Astronomie, Heidelberg. K.M. is also grateful to the IAU's Gruber Foundation Fellowship Programme for finanacial support. EC and PB to the list of people supported by ANR-18-CE31-0017. MB acknowledges the support to this research  by the PRIN INAF 2019 grant ObFu 1.05.01.85.14 (“Building up the halo: chemo-dynamical tagging in the age of large surveys”, PI. S. Lucatello). Y.H. is supported by National Key R\&D Program of China No. 2019YFA0405500 and National
Natural Science Foundation of China grants 11903027, 11833006, 11973001. ES acknowledges funding through VIDI grant "Pushing Galactic Archaeology to its limits" (with project number VI.Vidi.193.093) which is funded by the Dutch Research Council (NWO). DA acknowledges support from the ERC Starting Grant NEFERTITI H2020/808240.

We gratefully acknowledge the High Performance Computing center of the Université de Strasbourg for a very generous time allocation and for their support over the development of this project.

This work has made use of data from the European Space Agency (ESA) mission {\it Gaia} (\url{https://www.cosmos.esa.int/gaia}), processed by the {\it Gaia} Data Processing and Analysis Consortium (DPAC, \url{https://www.cosmos.esa.int/web/gaia/dpac/consortium}). Funding for the DPAC has been provided by national institutions, in particular the institutions participating in the {\it Gaia} Multilateral Agreement.

Based on observations collected at the European Southern Observatory under ESO programmes 0105.B-0235(A) and 0108.B-0431(A).

Based on data acquired at the Anglo-Australian Telescope. We acknowledge the traditional owners of the land on which the AAT stands, the Gamilaraay people, and pay our respects to elders past and present.

Software: STREAMFINDER \citep{malhan18a}, StarGO \citep{yuan18}, AGAMA \citep{agama}, astropy \citep{astropy}, galpy \citep{galpy}, IRAF \citep{tody86,tody93}, numpy \citet{numpy}, scipy \citep{scipy}, matplotlib \citep{matplotlib}, seaborn \citep{seaborn}.

\bibliography{ms}{}

\begin{thebibliography}{}
\expandafter\ifx\csname natexlab\endcsname\relax\def\natexlab#1{#1}\fi
\providecommand{\url}[1]{\href{#1}{#1}}
\providecommand{\dodoi}[1]{doi:~\href{http://doi.org/#1}{\nolinkurl{#1}}}
\providecommand{\doeprint}[1]{\href{http://ascl.net/#1}{\nolinkurl{http://ascl.net/#1}}}
\providecommand{\doarXiv}[1]{\href{https://arxiv.org/abs/#1}{\nolinkurl{https://arxiv.org/abs/#1}}}

\bibitem[{{Abbott} {et~al.}(2021){Abbott}, {Adam{\'o}w}, {Aguena}, {Allam},
  {Amon}, {Annis}, {Avila}, {Bacon}, {Banerji}, {Bechtol}, {Becker},
  {Bernstein}, {Bertin}, {Bhargava}, {Bridle}, {Brooks}, {Burke}, \& {Carnero
  Rosell}}]{desdr2}
{Abbott}, T.~M.~C., {Adam{\'o}w}, M., {Aguena}, M., {et~al.} 2021, \apjs, 255,
  20, \dodoi{10.3847/1538-4365/ac00b3}

\bibitem[{{Aguado} {et~al.}(2021{\natexlab{a}}){Aguado}, {Myeong}, {Belokurov},
  {Evans}, {Koposov}, {Allende Prieto}, {Lanfranchi}, {Matteucci}, {Shetrone},
  {Sbordone}, {Navarrete}, {Gonz{\'a}lez Hern{\'a}ndez}, {Chanam{\'e}},
  {Peralta de Arriba}, \& {Yuan}}]{aguado21a}
{Aguado}, D.~S., {Myeong}, G.~C., {Belokurov}, V., {et~al.} 2021{\natexlab{a}},
  \mnras, 500, 889, \dodoi{10.1093/mnras/staa3250}

\bibitem[{{Aguado} {et~al.}(2021{\natexlab{b}}){Aguado}, {Belokurov}, {Myeong},
  {Evans}, {Kobayashi}, {Sbordone}, {Chanam{\'e}}, {Navarrete}, \&
  {Koposov}}]{aguado21b}
{Aguado}, D.~S., {Belokurov}, V., {Myeong}, G.~C., {et~al.} 2021{\natexlab{b}},
  \apjl, 908, L8, \dodoi{10.3847/2041-8213/abdbb8}

\bibitem[{{Amorisco}(2017)}]{amorisco17}
{Amorisco}, N.~C. 2017, \mnras, 464, 2882, \dodoi{10.1093/mnras/stw2229}

\bibitem[{{Astropy Collaboration}(2018)}]{astropy}
{Astropy Collaboration}. 2018, \aj, 156, 123, \dodoi{10.3847/1538-3881/aabc4f}

\bibitem[{{Battaglia} {et~al.}(2021){Battaglia}, {Taibi}, {Thomas}, \&
  {Fritz}}]{battaglia21}
{Battaglia}, G., {Taibi}, S., {Thomas}, G.~F., \& {Fritz}, T.~K. 2021, arXiv
  e-prints, arXiv:2106.08819.
\newblock \doarXiv{2106.08819}

\bibitem[{{Baumgardt} \& {Makino}(2003)}]{baumgardt03}
{Baumgardt}, H., \& {Makino}, J. 2003, \mnras, 340, 227,
  \dodoi{10.1046/j.1365-8711.2003.06286.x}

\bibitem[{{Bechtol} {et~al.}(2015){Bechtol}, {Drlica-Wagner}, {Balbinot},
  {Pieres}, {Simon}, {Yanny}, {Santiago}, {Wechsler}, \& {Frieman}}]{bechtol15}
{Bechtol}, K., {Drlica-Wagner}, A., {Balbinot}, E., {et~al.} 2015, \apj, 807,
  50, \dodoi{10.1088/0004-637X/807/1/50}

\bibitem[{{Beers} \& {Christlieb}(2005)}]{beers05}
{Beers}, T.~C., \& {Christlieb}, N. 2005, \araa, 43, 531,
  \dodoi{10.1146/annurev.astro.42.053102.134057}

\bibitem[{{Bekki} \& {Freeman}(2003)}]{bekki03}
{Bekki}, K., \& {Freeman}, K.~C. 2003, \mnras, 346, L11,
  \dodoi{10.1046/j.1365-2966.2003.07275.x}

\bibitem[{{Bellazzini} {et~al.}(2020){Bellazzini}, {Ibata}, {Malhan}, {Martin},
  {Famaey}, \& {Thomas}}]{bellazzini20}
{Bellazzini}, M., {Ibata}, R., {Malhan}, K., {et~al.} 2020, \aap, 636, A107,
  \dodoi{10.1051/0004-6361/202037621}

\bibitem[{{Belokurov} {et~al.}(2018){Belokurov}, {Erkal}, {Evans}, {Koposov},
  \& {Deason}}]{belokurov18}
{Belokurov}, V., {Erkal}, D., {Evans}, N.~W., {Koposov}, S.~E., \& {Deason},
  A.~J. 2018, \mnras, 478, 611, \dodoi{10.1093/mnras/sty982}

\bibitem[{{Belokurov} {et~al.}(2006){Belokurov}, {Zucker}, {Evans},
  {Wilkinson}, {Irwin}, {Hodgkin}, {Bramich}, {Irwin}, {Gilmore}, {Willman},
  {Vidrih}, {Newberg}, {Wyse}, {Fellhauer}, {Hewett}, {Cole}, {Bell}, {Beers},
  {Rockosi}, {Yanny}, {Grebel}, {Schneider}, {Lupton}, {Barentine},
  {Brewington}, {Brinkmann}, {Harvanek}, {Kleinman}, {Krzesinski}, {Long},
  {Nitta}, {Smith}, \& {Snedden}}]{belokurov06}
{Belokurov}, V., {Zucker}, D.~B., {Evans}, N.~W., {et~al.} 2006, \apjl, 647,
  L111, \dodoi{10.1086/507324}

\bibitem[{{Bonaca} {et~al.}(2019){Bonaca}, {Conroy}, {Price-Whelan}, \&
  {Hogg}}]{bonaca19}
{Bonaca}, A., {Conroy}, C., {Price-Whelan}, A.~M., \& {Hogg}, D.~W. 2019,
  \apjl, 881, L37, \dodoi{10.3847/2041-8213/ab36ba}

\bibitem[{{Bonaca} {et~al.}(2012){Bonaca}, {Geha}, \&
  {Kallivayalil}}]{bonaca12}
{Bonaca}, A., {Geha}, M., \& {Kallivayalil}, N. 2012, \apjl, 760, L6,
  \dodoi{10.1088/2041-8205/760/1/L6}

\bibitem[{{Bonaca} {et~al.}(2021){Bonaca}, {Naidu}, {Conroy}, {Caldwell},
  {Cargile}, {Han}, {Johnson}, {Kruijssen}, {Myeong}, {Speagle}, {Ting}, \&
  {Zaritsky}}]{bonaca21}
{Bonaca}, A., {Naidu}, R.~P., {Conroy}, C., {et~al.} 2021, \apjl, 909, L26,
  \dodoi{10.3847/2041-8213/abeaa9}

\bibitem[{{Bovill} \& {Ricotti}(2009)}]{bovill09}
{Bovill}, M.~S., \& {Ricotti}, M. 2009, \apj, 693, 1859,
  \dodoi{10.1088/0004-637X/693/2/1859}

\bibitem[{{Bovy}(2015)}]{galpy}
{Bovy}, J. 2015, \apjs, 216, 29, \dodoi{10.1088/0067-0049/216/2/29}

\bibitem[{{Bressan} {et~al.}(2012){Bressan}, {Marigo}, {Girardi}, {Salasnich},
  {Dal Cero}, {Rubele}, \& {Nanni}}]{bressan12}
{Bressan}, A., {Marigo}, P., {Girardi}, L., {et~al.} 2012, \mnras, 427, 127,
  \dodoi{10.1111/j.1365-2966.2012.21948.x}

\bibitem[{{Brown} {et~al.}(2014){Brown}, {Tumlinson}, {Geha}, {Simon},
  {Vargas}, {VandenBerg}, {Kirby}, {Kalirai}, {Avila}, {Gennaro}, {Ferguson},
  {Mu{\~n}oz}, {Guhathakurta}, \& {Renzini}}]{brown14}
{Brown}, T.~M., {Tumlinson}, J., {Geha}, M., {et~al.} 2014, \apj, 796, 91,
  \dodoi{10.1088/0004-637X/796/2/91}

\bibitem[{{Buder} {et~al.}(2021){Buder}, {Lind}, {Ness}, \&
  {Feuillet}}]{buder21}
{Buder}, S., {Lind}, K., {Ness}, M.~K., \& {Feuillet}. 2021, arXiv e-prints,
  arXiv:2109.04059.
\newblock \doarXiv{2109.04059}

\bibitem[{{Bullock} \& {Boylan-Kolchin}(2017)}]{bullock17}
{Bullock}, J.~S., \& {Boylan-Kolchin}, M. 2017, \araa, 55, 343,
  \dodoi{10.1146/annurev-astro-091916-055313}

\bibitem[{{Bullock} \& {Johnston}(2005)}]{bullock05}
{Bullock}, J.~S., \& {Johnston}, K.~V. 2005, \apj, 635, 931,
  \dodoi{10.1086/497422}

\bibitem[{{Chambers} {et~al.}(2016){Chambers}, {Magnier}, {Metcalfe},
  {Flewelling}, {Huber}, {Waters}, {Denneau}, {Draper}, {Farrow}, {Finkbeiner},
  {Holmberg}, {Koppenhoefer}, {Price}, {Rest}, {Saglia}, {Schlafly}, {Smartt},
  {Sweeney}, {Wainscoat}, {Burgett}, {Chastel}, {Grav}, {Heasley}, \&
  {Hodapp}}]{chambers16}
{Chambers}, K.~C., {Magnier}, E.~A., {Metcalfe}, N., {et~al.} 2016, arXiv
  e-prints, arXiv:1612.05560.
\newblock \doarXiv{1612.05560}

\bibitem[{{Chang} {et~al.}(2020){Chang}, {Yuan}, {Xue}, {Simion}, {Kang}, {Li},
  {Zhao}, \& {Zhao}}]{chang20}
{Chang}, J., {Yuan}, Z., {Xue}, X.-X., {et~al.} 2020, \apj, 905, 100,
  \dodoi{10.3847/1538-4357/abc338}

\bibitem[{{Cui} {et~al.}(2012){Cui}, {Zhao}, {Chu}, {Li}, {Li}, {Zhang}, {Su},
  {Yao}, {Wang}, \& {Xing}}]{cui12}
{Cui}, X.-Q., {Zhao}, Y.-H., {Chu}, Y.-Q., {et~al.} 2012, Research in Astronomy
  and Astrophysics, 12, 1197, \dodoi{10.1088/1674-4527/12/9/003}

\bibitem[{{Deason} {et~al.}(2011){Deason}, {Belokurov}, \& {Evans}}]{deason11}
{Deason}, A.~J., {Belokurov}, V., \& {Evans}, N.~W. 2011, \mnras, 416, 2903,
  \dodoi{10.1111/j.1365-2966.2011.19237.x}

\bibitem[{{Dierickx} \& {Loeb}(2017)}]{dierickx17}
{Dierickx}, M. I.~P., \& {Loeb}, A. 2017, \apj, 836, 92,
  \dodoi{10.3847/1538-4357/836/1/92}

\bibitem[{{Donlon} {et~al.}(2020){Donlon}, {Newberg}, {Sanderson}, \&
  {Widrow}}]{donlon20}
{Donlon}, Thomas, I., {Newberg}, H.~J., {Sanderson}, R., \& {Widrow}, L.~M.
  2020, \apj, 902, 119, \dodoi{10.3847/1538-4357/abb5f6}

\bibitem[{{Drlica-Wagner} {et~al.}(2015){Drlica-Wagner}, {Bechtol}, {Rykoff},
  \& {DES Collaboration}}]{wagner15}
{Drlica-Wagner}, A., {Bechtol}, K., {Rykoff}, E.~S., \& {DES Collaboration}.
  2015, \apj, 813, 109, \dodoi{10.1088/0004-637X/813/2/109}

\bibitem[{{Duch} \& {Naud}(1996)}]{lodzis}
{Duch}, W., \& {Naud}, A. 1996, Proceedings of the Second Conference on Neural
  Networks and their applications, 30.IV-4.V., 138

\bibitem[{{Gaia Collaboration} {et~al.}(2018){Gaia Collaboration}, {Brown},
  {Vallenari}, \& {Prusti}}]{brown18}
{Gaia Collaboration}, {Brown}, A.~G.~A., {Vallenari}, A., \& {Prusti}. 2018,
  \aap, 616, A1, \dodoi{10.1051/0004-6361/201833051}

\bibitem[{{Gaia Collaboration} {et~al.}(2016){Gaia Collaboration}, {Prusti},
  {de Bruijne}, {Brown}, \& {Vallenari}}]{gaia}
{Gaia Collaboration}, {Prusti}, T., {de Bruijne}, J.~H.~J., {Brown}, A.~G.~A.,
  \& {Vallenari}. 2016, \aap, 595, A1, \dodoi{10.1051/0004-6361/201629272}

\bibitem[{{Georgiev} {et~al.}(2016){Georgiev}, {B{\"o}ker}, {Leigh},
  {L{\"u}tzgendorf}, \& {Neumayer}}]{georgiev16}
{Georgiev}, I.~Y., {B{\"o}ker}, T., {Leigh}, N., {L{\"u}tzgendorf}, N., \&
  {Neumayer}, N. 2016, \mnras, 457, 2122, \dodoi{10.1093/mnras/stw093}

\bibitem[{{Gudin} {et~al.}(2021){Gudin}, {Shank}, {Beers}, {Yuan}, {Limberg},
  {Roederer}, {Placco}, {Holmbeck}, {Dietz}, {Rasmussen}, {Hansen}, {Sakari},
  {Ezzeddine}, \& {Frebel}}]{gudin21}
{Gudin}, D., {Shank}, D., {Beers}, T.~C., {et~al.} 2021, \apj, 908, 79,
  \dodoi{10.3847/1538-4357/abd7ed}

\bibitem[{{Gull} {et~al.}(2021){Gull}, {Frebel}, {Hinojosa}, {Roederer}, {Ji},
  \& {Brauer}}]{gull21}
{Gull}, M., {Frebel}, A., {Hinojosa}, K., {et~al.} 2021, \apj, 912, 52,
  \dodoi{10.3847/1538-4357/abea1a}

\bibitem[{{Harris} {et~al.}(2013){Harris}, {Harris}, \& {Alessi}}]{harris13}
{Harris}, W.~E., {Harris}, G. L.~H., \& {Alessi}, M. 2013, \apj, 772, 82,
  \dodoi{10.1088/0004-637X/772/2/82}

\bibitem[{{Helmi} {et~al.}(2018){Helmi}, {Babusiaux}, {Koppelman}, {Massari},
  {Veljanoski}, \& {Brown}}]{helmi18}
{Helmi}, A., {Babusiaux}, C., {Koppelman}, H.~H., {et~al.} 2018, \nat, 563, 85,
  \dodoi{10.1038/s41586-018-0625-x}

\bibitem[{{Helmi} {et~al.}(1999){Helmi}, {White}, {de Zeeuw}, \&
  {Zhao}}]{helmi99}
{Helmi}, A., {White}, S. D.~M., {de Zeeuw}, P.~T., \& {Zhao}, H. 1999, \nat,
  402, 53, \dodoi{10.1038/46980}

\bibitem[{{Huang} {et~al.}(2021{\natexlab{a}}){Huang}, {Yuan}, {Li}, {Wolf},
  {Onken}, {Beers}, {Casagrande}, {Mackey}, {Da Costa}, {Bland-Hawthorn},
  {Stello}, {Nordlander}, {Ting}, {Buder}, {Sharma}, \& {Liu}}]{huang21a}
{Huang}, Y., {Yuan}, H., {Li}, C., {et~al.} 2021{\natexlab{a}}, \apj, 907, 68,
  \dodoi{10.3847/1538-4357/abca37}

\bibitem[{{Huang} {et~al.}(2021{\natexlab{b}}){Huang}, {Beers}, {Wolf}, {Lee},
  {Onken}, {Yuan}, {Shank}, {Zhang}, {Wang}, {Shi}, \& {Fan}}]{huang21b}
{Huang}, Y., {Beers}, T.~C., {Wolf}, C., {et~al.} 2021{\natexlab{b}}, arXiv
  e-prints, arXiv:2104.14154.
\newblock \doarXiv{2104.14154}

\bibitem[{{Hunter}(2007)}]{matplotlib}
{Hunter}, J.~D. 2007, Computing in Science and Engineering, 9, 90,
  \dodoi{10.1109/MCSE.2007.55}

\bibitem[{{Ibata} {et~al.}(2001){Ibata}, {Irwin}, {Lewis}, \&
  {Stolte}}]{ibata01}
{Ibata}, R., {Irwin}, M., {Lewis}, G.~F., \& {Stolte}, A. 2001, \apjl, 547,
  L133, \dodoi{10.1086/318894}

\bibitem[{{Ibata} {et~al.}(2021){Ibata}, {Malhan}, {Martin}, {Aubert},
  {Famaey}, {Bianchini}, {Monari}, {Siebert}, {Thomas}, {Bellazzini},
  {Bonifacio}, {Caffau}, \& {Renaud}}]{ibata21}
{Ibata}, R., {Malhan}, K., {Martin}, N., {et~al.} 2021, \apj, 914, 123,
  \dodoi{10.3847/1538-4357/abfcc2}

\bibitem[{{Ibata} {et~al.}(2019{\natexlab{a}}){Ibata}, {Bellazzini}, {Malhan},
  {Martin}, \& {Bianchini}}]{ibata19b}
{Ibata}, R.~A., {Bellazzini}, M., {Malhan}, K., {Martin}, N., \& {Bianchini},
  P. 2019{\natexlab{a}}, Nature Astronomy, 3, 667,
  \dodoi{10.1038/s41550-019-0751-x}

\bibitem[{{Ibata} {et~al.}(2019{\natexlab{b}}){Ibata}, {Malhan}, \&
  {Martin}}]{ibata19a}
{Ibata}, R.~A., {Malhan}, K., \& {Martin}, N.~F. 2019{\natexlab{b}}, \apj, 872,
  152, \dodoi{10.3847/1538-4357/ab0080}

\bibitem[{{Ibata} {et~al.}(2018){Ibata}, {Malhan}, {Martin}, \&
  {Starkenburg}}]{ibata18}
{Ibata}, R.~A., {Malhan}, K., {Martin}, N.~F., \& {Starkenburg}, E. 2018, \apj,
  865, 85, \dodoi{10.3847/1538-4357/aadba3}

\bibitem[{{Ideta} \& {Makino}(2004)}]{ideta04}
{Ideta}, M., \& {Makino}, J. 2004, \apjl, 616, L107, \dodoi{10.1086/426505}

\bibitem[{{Irwin} {et~al.}(2007){Irwin}, {Belokurov}, {Evans}, {Ryan-Weber},
  {de Jong}, {Koposov}, {Zucker}, {Hodgkin}, {Gilmore}, {Prema}, {Hebb},
  {Begum}, {Fellhauer}, {Hewett}, {Kennicutt}, {Wilkinson}, {Bramich},
  {Vidrih}, {Rix}, {Beers}, {Barentine}, {Brewington}, {Harvanek},
  {Krzesinski}, {Long}, {Nitta}, \& {Snedden}}]{irwin07}
{Irwin}, M.~J., {Belokurov}, V., {Evans}, N.~W., {et~al.} 2007, \apjl, 656,
  L13, \dodoi{10.1086/512183}

\bibitem[{{Ji} {et~al.}(2020){Ji}, {Li}, {Hansen}, {Casey}, {Koposov}, {Pace},
  {Mackey}, {Lewis}, {Simpson}, {Bland-Hawthorn}, {Cullinane}, {Da Costa},
  {Hattori}, {Martell}, {Kuehn}, {Erkal}, {Shipp}, {Wan}, \& {Zucker}}]{ji20}
{Ji}, A.~P., {Li}, T.~S., {Hansen}, T.~T., {et~al.} 2020, \aj, 160, 181,
  \dodoi{10.3847/1538-3881/abacb6}

\bibitem[{{Johnston}(1998)}]{johnston98}
{Johnston}, K.~V. 1998, \apj, 495, 297, \dodoi{10.1086/305273}

\bibitem[{Jones {et~al.}(2001--)Jones, Oliphant, Peterson, {et~al.}}]{scipy}
Jones, E., Oliphant, T., Peterson, P., {et~al.} 2001--, {SciPy}: Open source
  scientific tools for {Python}.
\newblock \url{http://www.scipy.org/}

\bibitem[{{Kirby} {et~al.}(2013){Kirby}, {Cohen}, {Guhathakurta}, {Cheng},
  {Bullock}, \& {Gallazzi}}]{kirby13}
{Kirby}, E.~N., {Cohen}, J.~G., {Guhathakurta}, P., {et~al.} 2013, \apj, 779,
  102, \dodoi{10.1088/0004-637X/779/2/102}

\bibitem[{{Kohonen }(1982)}]{kohonen82}
{Kohonen }, T. 1982, Biological Cybernetics, 43, 59,
  \dodoi{https://doi.org/10.1007/BF00337288}

\bibitem[{{Koposov} {et~al.}(2008){Koposov}, {Belokurov}, {Evans}, {Hewett},
  {Irwin}, {Gilmore}, {Zucker}, {Rix}, {Fellhauer}, {Bell}, \&
  {Glushkova}}]{koposov08}
{Koposov}, S., {Belokurov}, V., {Evans}, N.~W., {et~al.} 2008, \apj, 686, 279,
  \dodoi{10.1086/589911}

\bibitem[{{Koposov} {et~al.}(2015){Koposov}, {Belokurov}, {Torrealba}, \&
  {Evans}}]{koposov15}
{Koposov}, S.~E., {Belokurov}, V., {Torrealba}, G., \& {Evans}, N.~W. 2015,
  \apj, 805, 130, \dodoi{10.1088/0004-637X/805/2/130}

\bibitem[{{Koppelman} {et~al.}(2019){Koppelman}, {Helmi}, {Massari},
  {Price-Whelan}, \& {Starkenburg}}]{koppelman19}
{Koppelman}, H.~H., {Helmi}, A., {Massari}, D., {Price-Whelan}, A.~M., \&
  {Starkenburg}, T.~K. 2019, \aap, 631, L9, \dodoi{10.1051/0004-6361/201936738}

\bibitem[{{Kroupa}(2001)}]{kroupa21}
{Kroupa}, P. 2001, \mnras, 322, 231, \dodoi{10.1046/j.1365-8711.2001.04022.x}

\bibitem[{{Kroupa}(2002)}]{kroupa22}
---. 2002, Science, 295, 82, \dodoi{10.1126/science.1067524}

\bibitem[{{Kruijssen}(2019)}]{kruijssen19}
{Kruijssen}, J.~M.~D. 2019, \mnras, 486, L20, \dodoi{10.1093/mnrasl/slz052}

\bibitem[{{Laevens} {et~al.}(2015){Laevens}, {Martin}, {Bernard}, {Schlafly},
  {Sesar}, {Rix}, {Bell}, {Ferguson}, {Slater}, {Sweeney}, {Wyse}, {Huxor},
  {Burgett}, {Chambers}, {Draper}, {Hodapp}, {Kaiser}, {Magnier}, {Metcalfe},
  {Tonry}, {Wainscoat}, \& {Waters}}]{laevens15b}
{Laevens}, B. P.~M., {Martin}, N.~F., {Bernard}, E.~J., {et~al.} 2015, \apj,
  813, 44, \dodoi{10.1088/0004-637X/813/1/44}

\bibitem[{{Law} \& {Majewski}(2010)}]{law10}
{Law}, D.~R., \& {Majewski}, S.~R. 2010, \apj, 714, 229,
  \dodoi{10.1088/0004-637X/714/1/229}

\bibitem[{{Li} {et~al.}(2018){Li}, {Tan}, \& {Zhao}}]{li18}
{Li}, H., {Tan}, K., \& {Zhao}, G. 2018, \apjs, 238, 16,
  \dodoi{10.3847/1538-4365/aada4a}

\bibitem[{{Li} {et~al.}(2016){Li}, {Balbinot}, {Mondrik}, {Marshall}, {Yanny},
  {Bechtol}, \& {Drlica-Wagner}}]{li16}
{Li}, T.~S., {Balbinot}, E., {Mondrik}, N., {et~al.} 2016, \apj, 817, 135,
  \dodoi{10.3847/0004-637X/817/2/135}

\bibitem[{{Li} {et~al.}(2019){Li}, {Koposov}, {Zucker}, {Lewis}, {Kuehn},
  {Simpson}, {Ji}, {Shipp}, {Mao}, {Geha}, {Pace}, {Mackey}, {Allam}, {Tucker},
  {Da Costa}, {Erkal}, {Simon}, {Mould}, {Martell}, {Wan}, {De Silva},
  {Bechtol}, {Balbinot}, {Belokurov}, {Bland-Hawthorn}, {Casey}, {Cullinane},
  {Drlica-Wagner}, {Sharma}, {Vivas}, {Wechsler}, {Yanny}, \& {S5
  Collaboration}}]{li19}
{Li}, T.~S., {Koposov}, S.~E., {Zucker}, D.~B., {et~al.} 2019, \mnras, 490,
  3508, \dodoi{10.1093/mnras/stz2731}

\bibitem[{{Li} {et~al.}(2021{\natexlab{a}}){Li}, {Ji}, {Pace}, {Erkal},
  {Koposov}, {Shipp}, {Da Costa}, {Cullinane}, {Kuehn}, {Lewis}, {Mackey},
  {Simpson}, {Zucker}, {Ferguson}, {Martell}, {Bland-Hawthorn}, {Balbinot},
  {Tavangar}, {Drlica-Wagner}, {De Silva1}, {Simon}, \& {S5
  Collaboration}}]{li21b}
{Li}, T.~S., {Ji}, A.~P., {Pace}, A.~B., {et~al.} 2021{\natexlab{a}}, arXiv
  e-prints, arXiv:2110.06950.
\newblock \doarXiv{2110.06950}

\bibitem[{{Li} {et~al.}(2021{\natexlab{b}}){Li}, {Koposov}, {Erkal}, {Ji},
  {Shipp}, {Pace}, {Hilmi}, {Kuehn}, \& {Lewis}}]{li21a}
{Li}, T.~S., {Koposov}, S.~E., {Erkal}, D., {et~al.} 2021{\natexlab{b}}, \apj,
  911, 149, \dodoi{10.3847/1538-4357/abeb18}

\bibitem[{{Limberg} {et~al.}(2021){Limberg}, {Rossi}, {Beers}, {Perottoni},
  {P{\'e}rez-Villegas}, {Santucci}, {Abuchaim}, {Placco}, {Lee}, {Christlieb},
  {Norris}, {Bessell}, {Ryan}, {Wilhelm}, {Rhee}, \& {Frebel}}]{limberg21}
{Limberg}, G., {Rossi}, S., {Beers}, T.~C., {et~al.} 2021, \apj, 907, 10,
  \dodoi{10.3847/1538-4357/abcb87}

\bibitem[{{Majewski} {et~al.}(2003){Majewski}, {Skrutskie}, {Weinberg}, \&
  {Ostheimer}}]{majewski03}
{Majewski}, S.~R., {Skrutskie}, M.~F., {Weinberg}, M.~D., \& {Ostheimer}, J.~C.
  2003, \apj, 599, 1082, \dodoi{10.1086/379504}

\bibitem[{{Malhan} \& {Ibata}(2018)}]{malhan18a}
{Malhan}, K., \& {Ibata}, R.~A. 2018, \mnras, 477, 4063,
  \dodoi{10.1093/mnras/sty912}

\bibitem[{{Malhan} {et~al.}(2018){Malhan}, {Ibata}, \& {Martin}}]{malhan18b}
{Malhan}, K., {Ibata}, R.~A., \& {Martin}, N.~F. 2018, \mnras, 481, 3442,
  \dodoi{10.1093/mnras/sty2474}

\bibitem[{{Malhan} {et~al.}(2021){Malhan}, {Yuan}, {Ibata}, {Arentsen},
  {Bellazzini}, \& {Martin}}]{malhan21}
{Malhan}, K., {Yuan}, Z., {Ibata}, R.~A., {et~al.} 2021, \apj, 920, 51,
  \dodoi{10.3847/1538-4357/ac1675}

\bibitem[{{Martin} {et~al.}(2013){Martin}, {Carlin}, {Newberg}, \&
  {Grillmair}}]{martin13}
{Martin}, C., {Carlin}, J.~L., {Newberg}, H.~J., \& {Grillmair}, C. 2013,
  \apjl, 765, L39, \dodoi{10.1088/2041-8205/765/2/L39}

\bibitem[{{Martin} {et~al.}(2018){Martin}, {Collins}, {Longeard}, \&
  {Tollerud}}]{martin18}
{Martin}, N.~F., {Collins}, M. L.~M., {Longeard}, N., \& {Tollerud}, E. 2018,
  \apjl, 859, L5, \dodoi{10.3847/2041-8213/aac216}

\bibitem[{{Martin} {et~al.}(2014){Martin}, {Ibata}, {Rich}, \&
  {Collins}}]{martin14}
{Martin}, N.~F., {Ibata}, R.~A., {Rich}, R.~M., \& {Collins}. 2014, \apj, 787,
  19, \dodoi{10.1088/0004-637X/787/1/19}

\bibitem[{{Martin} {et~al.}(2015){Martin}, {Nidever}, {Besla}, \&
  {Olsen}}]{martin15}
{Martin}, N.~F., {Nidever}, D.~L., {Besla}, G., \& {Olsen}. 2015, \apjl, 804,
  L5, \dodoi{10.1088/2041-8205/804/1/L5}

\bibitem[{{Martin} {et~al.}(2022{\natexlab{a}}){Martin}, {Venn}, {Aguado},
  {Starkenburg}, {Gonz{\'a}lez Hern{\'a}ndez}, {Ibata}, {Bonifacio}, {Caffau},
  {Sestito}, {Arentsen}, {Allende Prieto}, {Carlberg}, {Fabbro}, {Fouesneau},
  {Hill}, {Jablonka}, {Kordopatis}, {Lardo}, {Malhan}, {Mashonkina},
  {McConnachie}, {Navarro}, {S{\'a}nchez-Janssen}, {Thomas}, {Yuan}, \&
  {Mucciarelli}}]{martin22a}
{Martin}, N.~F., {Venn}, K.~A., {Aguado}, D.~S., {et~al.} 2022{\natexlab{a}},
  \nat, 601, 45, \dodoi{10.1038/s41586-021-04162-2}

\bibitem[{{Martin} {et~al.}(2022{\natexlab{b}}){Martin}, {Ibata},
  {Starkenburg}, {Yuan}, {Malhan}, {Bellazzini}, {Viswanathan}, {Aguado},
  {Arentsen}, {Bonifacio}, {Carlberg}, {Gonz{\'a}lez Hern{\'a}ndez}, {Hill},
  {Jablonka}, {Kordopatis}, {Lardo}, {McConnachie}, {Navarro},
  {S{\'a}nchez-Janssen}, {Sestito}, {Thomas}, {Venn}, {Vitali}, \&
  {Voggel}}]{martin22b}
{Martin}, N.~F., {Ibata}, R.~A., {Starkenburg}, E., {et~al.}
  2022{\natexlab{b}}, arXiv e-prints, arXiv:2201.01310.
\newblock \doarXiv{2201.01310}

\bibitem[{{Massari} {et~al.}(2019){Massari}, {Koppelman}, \&
  {Helmi}}]{massari19}
{Massari}, D., {Koppelman}, H.~H., \& {Helmi}, A. 2019, \aap, 630, L4,
  \dodoi{10.1051/0004-6361/201936135}

\bibitem[{{Mateo} {et~al.}(1996){Mateo}, {Mirabal}, {Udalski}, {Szymanski},
  {Kaluzny}, {Kubiak}, {Krzeminski}, \& {Stanek}}]{mateo96}
{Mateo}, M., {Mirabal}, N., {Udalski}, A., {et~al.} 1996, \apjl, 458, L13,
  \dodoi{10.1086/309919}

\bibitem[{{Matsuno} {et~al.}(2019){Matsuno}, {Aoki}, \& {Suda}}]{matsuno19}
{Matsuno}, T., {Aoki}, W., \& {Suda}, T. 2019, \apjl, 874, L35,
  \dodoi{10.3847/2041-8213/ab0ec0}

\bibitem[{{Matsuno} {et~al.}(2021){Matsuno}, {Hirai}, {Tarumi}, {Hotokezaka},
  {Tanaka}, \& {Helmi}}]{matsuno21}
{Matsuno}, T., {Hirai}, Y., {Tarumi}, Y., {et~al.} 2021, \aap, 650, A110,
  \dodoi{10.1051/0004-6361/202040227}

\bibitem[{{McMillan}(2017)}]{mc17}
{McMillan}, P.~J. 2017, \mnras, 465, 76, \dodoi{10.1093/mnras/stw2759}

\bibitem[{{Mizutani} {et~al.}(2003){Mizutani}, {Chiba}, \&
  {Sakamoto}}]{mizutani03}
{Mizutani}, A., {Chiba}, M., \& {Sakamoto}, T. 2003, \apjl, 589, L89,
  \dodoi{10.1086/375873}

\bibitem[{{Monachesi} {et~al.}(2019){Monachesi}, {G{\'o}mez}, {Grand},
  {Simpson}, {Kauffmann}, {Bustamante}, {Marinacci}, {Pakmor}, {Springel},
  {Frenk}, {White}, \& {Tissera}}]{monachesi19}
{Monachesi}, A., {G{\'o}mez}, F.~A., {Grand}, R. J.~J., {et~al.} 2019, \mnras,
  485, 2589, \dodoi{10.1093/mnras/stz538}

\bibitem[{{Mucciarelli} {et~al.}(2018){Mucciarelli}, {Lapenna}, {Ferraro}, \&
  {Lanzoni}}]{mucciarelli18}
{Mucciarelli}, A., {Lapenna}, E., {Ferraro}, F.~R., \& {Lanzoni}, B. 2018,
  \apj, 859, 75, \dodoi{10.3847/1538-4357/aaba80}

\bibitem[{{Myeong} {et~al.}(2018{\natexlab{a}}){Myeong}, {Evans}, {Belokurov},
  {Amorisco}, \& {Koposov}}]{myeong18a}
{Myeong}, G.~C., {Evans}, N.~W., {Belokurov}, V., {Amorisco}, N.~C., \&
  {Koposov}, S.~E. 2018{\natexlab{a}}, \mnras, 475, 1537,
  \dodoi{10.1093/mnras/stx3262}

\bibitem[{{Myeong} {et~al.}(2018{\natexlab{b}}){Myeong}, {Evans}, {Belokurov},
  {Sanders}, \& {Koposov}}]{myeong18b}
{Myeong}, G.~C., {Evans}, N.~W., {Belokurov}, V., {Sanders}, J.~L., \&
  {Koposov}, S.~E. 2018{\natexlab{b}}, \apjl, 863, L28,
  \dodoi{10.3847/2041-8213/aad7f7}

\bibitem[{{Myeong} {et~al.}(2019){Myeong}, {Vasiliev}, {Iorio}, {Evans}, \&
  {Belokurov}}]{myeong19}
{Myeong}, G.~C., {Vasiliev}, E., {Iorio}, G., {Evans}, N.~W., \& {Belokurov},
  V. 2019, \mnras, 488, 1235, \dodoi{10.1093/mnras/stz1770}

\bibitem[{{Naidu} {et~al.}(2020){Naidu}, {Conroy}, {Bonaca}, \&
  {Johnson}}]{naidu20}
{Naidu}, R.~P., {Conroy}, C., {Bonaca}, A., \& {Johnson}. 2020, \apj, 901, 48,
  \dodoi{10.3847/1538-4357/abaef4}

\bibitem[{{Newberg} {et~al.}(2009){Newberg}, {Yanny}, \& {Willett}}]{newberg09}
{Newberg}, H.~J., {Yanny}, B., \& {Willett}, B.~A. 2009, \apjl, 700, L61,
  \dodoi{10.1088/0004-637X/700/2/L61}

\bibitem[{{Pe{\~n}arrubia} {et~al.}(2010){Pe{\~n}arrubia}, {Belokurov},
  {Evans}, {Mart{\'\i}nez-Delgado}, {Gilmore}, {Irwin}, {Niederste-Ostholt}, \&
  {Zucker}}]{penarrubia10}
{Pe{\~n}arrubia}, J., {Belokurov}, V., {Evans}, N.~W., {et~al.} 2010, \mnras,
  408, L26, \dodoi{10.1111/j.1745-3933.2010.00921.x}

\bibitem[{{Price-Whelan} \& {Bonaca}(2018)}]{price18}
{Price-Whelan}, A.~M., \& {Bonaca}, A. 2018, \apjl, 863, L20,
  \dodoi{10.3847/2041-8213/aad7b5}

\bibitem[{{Read} {et~al.}(2017){Read}, {Iorio}, {Agertz}, \&
  {Fraternali}}]{read17}
{Read}, J.~I., {Iorio}, G., {Agertz}, O., \& {Fraternali}, F. 2017, \mnras,
  467, 2019, \dodoi{10.1093/mnras/stx147}

\bibitem[{{Riello} {et~al.}(2021){Riello}, {De Angeli}, \& {Evans}}]{riello21}
{Riello}, M., {De Angeli}, F., \& {Evans}, D.~W. 2021, \aap, 649, A3,
  \dodoi{10.1051/0004-6361/202039587}

\bibitem[{{Roederer} {et~al.}(2018){Roederer}, {Hattori}, \&
  {Valluri}}]{roederer18}
{Roederer}, I.~U., {Hattori}, K., \& {Valluri}, M. 2018, \aj, 156, 179,
  \dodoi{10.3847/1538-3881/aadd9c}

\bibitem[{{Roederer} {et~al.}(2016){Roederer}, {Mateo}, {Bailey}, {Song},
  {Bell}, {Crane}, {Loebman}, {Nidever}, {Olszewski}, {Shectman}, {Thompson},
  {Valluri}, \& {Walker}}]{roederer16}
{Roederer}, I.~U., {Mateo}, M., {Bailey}, John~I., I., {et~al.} 2016, \aj, 151,
  82, \dodoi{10.3847/0004-6256/151/3/82}

\bibitem[{{Sbordone} {et~al.}(2014){Sbordone}, {Caffau}, {Bonifacio}, \&
  {Duffau}}]{sbordone14}
{Sbordone}, L., {Caffau}, E., {Bonifacio}, P., \& {Duffau}, S. 2014, \aap, 564,
  A109, \dodoi{10.1051/0004-6361/201322430}

\bibitem[{{Schlafly} \& {Finkbeiner}(2011)}]{schlafly11}
{Schlafly}, E.~F., \& {Finkbeiner}, D.~P. 2011, \apj, 737, 103,
  \dodoi{10.1088/0004-637X/737/2/103}

\bibitem[{{Schlegel} {et~al.}(1998){Schlegel}, {Finkbeiner}, \&
  {Davis}}]{schlegel98}
{Schlegel}, D.~J., {Finkbeiner}, D.~P., \& {Davis}, M. 1998, \apj, 500, 525,
  \dodoi{10.1086/305772}

\bibitem[{{Sestito} {et~al.}(2020){Sestito}, {Martin}, {Starkenburg}, \&
  {Arentsen}}]{sestito20}
{Sestito}, F., {Martin}, N.~F., {Starkenburg}, E., \& {Arentsen}. 2020, \mnras,
  497, L7, \dodoi{10.1093/mnrasl/slaa022}

\bibitem[{{Sestito} {et~al.}(2019){Sestito}, {Longeard}, {Martin},
  {Starkenburg}, {Fouesneau}, {Gonz{\'a}lez Hern{\'a}ndez}, {Arentsen},
  {Ibata}, {Aguado}, {Carlberg}, {Jablonka}, {Navarro}, {Tolstoy}, \&
  {Venn}}]{sestito19}
{Sestito}, F., {Longeard}, N., {Martin}, N.~F., {et~al.} 2019, \mnras, 484,
  2166, \dodoi{10.1093/mnras/stz043}

\bibitem[{{Shank} {et~al.}(2021){Shank}, {Beers}, {Placco}, {Limberg},
  {Jaques}, {Yuan}, {Schlaufman}, {Casey}, {Huang}, {Lee}, {Hattori}, \&
  {Santucci}}]{shank21}
{Shank}, D., {Beers}, T.~C., {Placco}, V.~M., {et~al.} 2021, arXiv e-prints,
  arXiv:2109.08600.
\newblock \doarXiv{2109.08600}

\bibitem[{{Shipp} {et~al.}(2018){Shipp}, {Drlica-Wagner}, {Balbinot}, \&
  {Ferguson}}]{shipp18}
{Shipp}, N., {Drlica-Wagner}, A., {Balbinot}, E., \& {Ferguson}, P. 2018, \apj,
  862, 114, \dodoi{10.3847/1538-4357/aacdab}

\bibitem[{{Simon}(2019)}]{simon19}
{Simon}, J.~D. 2019, \araa, 57, 375,
  \dodoi{10.1146/annurev-astro-091918-104453}

\bibitem[{{Simpson} {et~al.}(2021){Simpson}, {Martell}, {Buder},
  {Bland-Hawthorn}, {Casey}, {de Silva}, {D'Orazi}, {Freeman}, {Hayden}, {Kos},
  {Lewis}, {Lind}, {Schlesinger}, {Sharma}, {Stello}, {Zucker}, {Zwitter},
  {Asplund}, {da Costa}, {{\v{C}}otar}, {Tepper-Garc{\'\i}a}, {Horner},
  {Nordlander}, {Ting}, {Wyse}, \& {GALAH Collaboration}}]{simpson21}
{Simpson}, J.~D., {Martell}, S.~L., {Buder}, S., {et~al.} 2021, \mnras, 507,
  43, \dodoi{10.1093/mnras/stab2012}

\bibitem[{{Starkenburg} {et~al.}(2017){Starkenburg}, {Martin}, {Youakim},
  {Aguado}, {Allende Prieto}, {Arentsen}, {Bernard}, {Bonifacio}, {Caffau},
  {Carlberg}, {C{\^o}t{\'e}}, {Fouesneau}, {Fran{\c{c}}ois}, {Franke},
  {Gonz{\'a}lez Hern{\'a}ndez}, {Gwyn}, {Hill}, {Ibata}, {Jablonka},
  {Longeard}, {McConnachie}, {Navarro}, {S{\'a}nchez-Janssen}, {Tolstoy}, \&
  {Venn}}]{starkenburg17}
{Starkenburg}, E., {Martin}, N., {Youakim}, K., {et~al.} 2017, \mnras, 471,
  2587, \dodoi{10.1093/mnras/stx1068}

\bibitem[{{The Dark Energy Survey Collaboration}(2005)}]{des}
{The Dark Energy Survey Collaboration}. 2005, arXiv e-prints, astro.
\newblock \doarXiv{astro-ph/0510346}

\bibitem[{{Tody}(1986)}]{tody86}
{Tody}, D. 1986, in Society of Photo-Optical Instrumentation Engineers (SPIE)
  Conference Series, Vol. 627, Instrumentation in astronomy VI, ed. D.~L.
  {Crawford}, 733, \dodoi{10.1117/12.968154}

\bibitem[{{Tody}(1993)}]{tody93}
{Tody}, D. 1993, in Astronomical Society of the Pacific Conference Series,
  Vol.~52, Astronomical Data Analysis Software and Systems II, ed. R.~J.
  {Hanisch}, R.~J.~V. {Brissenden}, \& J.~{Barnes}, 173

\bibitem[{{Tsuchiya} {et~al.}(2004){Tsuchiya}, {Korchagin}, \&
  {Dinescu}}]{tsuchiya04}
{Tsuchiya}, T., {Korchagin}, V.~I., \& {Dinescu}, D.~I. 2004, \mnras, 350,
  1141, \dodoi{10.1111/j.1365-2966.2004.07716.x}

\bibitem[{{van der Walt} {et~al.}(2011){van der Walt}, {Colbert}, \&
  {Varoquaux}}]{numpy}
{van der Walt}, S., {Colbert}, S.~C., \& {Varoquaux}, G. 2011, Computing in
  Science and Engineering, 13, 22, \dodoi{10.1109/MCSE.2011.37}

\bibitem[{{Vasiliev}(2019)}]{agama}
{Vasiliev}, E. 2019, \mnras, 482, 1525, \dodoi{10.1093/mnras/sty2672}

\bibitem[{{Vasiliev} \& {Baumgardt}(2021)}]{vasiliev21}
{Vasiliev}, E., \& {Baumgardt}, H. 2021, \mnras, 505, 5978,
  \dodoi{10.1093/mnras/stab1475}

\bibitem[{{Vasiliev} {et~al.}(2021){Vasiliev}, {Belokurov}, \&
  {Erkal}}]{vasiliev21b}
{Vasiliev}, E., {Belokurov}, V., \& {Erkal}, D. 2021, \mnras, 501, 2279,
  \dodoi{10.1093/mnras/staa3673}

\bibitem[{{Vera-Ciro} \& {Helmi}(2013)}]{vera13}
{Vera-Ciro}, C., \& {Helmi}, A. 2013, \apjl, 773, L4,
  \dodoi{10.1088/2041-8205/773/1/L4}

\bibitem[{{Wan} {et~al.}(2020){Wan}, {Lewis}, {Li}, {Simpson}, {Martell},
  {Zucker}, {Mould}, {Erkal}, {Pace}, {Mackey}, {Ji}, {Koposov}, {Kuehn},
  {Shipp}, {Balbinot}, {Bland-Hawthorn}, {Casey}, {Da Costa}, {Kafle},
  {Sharma}, \& {De Silva}}]{wan20}
{Wan}, Z., {Lewis}, G.~F., {Li}, T.~S., {et~al.} 2020, \nat, 583, 768,
  \dodoi{10.1038/s41586-020-2483-6}

\bibitem[{{Waskom} {et~al.}(2016){Waskom}, {Botvinnik}, {drewokane}, {Hobson},
  {Halchenko}, {Lukauskas}, {Warmenhoven}, {Cole}, {Hoyer}, {Vanderplas},
  {gkunter}, {Villalba}, {Quintero}, {Martin}, {Miles}, {Meyer}, {Augspurger},
  {Yarkoni}, {Bachant}, {Evans}, {Fitzgerald}, {Nagy}, {Ziegler}, {Megies},
  {Wehner}, {St-Jean}, {Coelho}, {Hitz}, {Lee}, \& {Rocher}}]{seaborn}
{Waskom}, M., {Botvinnik}, O., {drewokane}, {et~al.} 2016, {Seaborn: V0.7.0
  (January 2016)}, v0.7.0,  Zenodo, \dodoi{10.5281/zenodo.45133}

\bibitem[{{Willman} {et~al.}(2005{\natexlab{a}}){Willman}, {Blanton}, {West},
  {Dalcanton}, {Hogg}, {Schneider}, {Wherry}, {Yanny}, \&
  {Brinkmann}}]{willman05a}
{Willman}, B., {Blanton}, M.~R., {West}, A.~A., {et~al.} 2005{\natexlab{a}},
  \aj, 129, 2692, \dodoi{10.1086/430214}

\bibitem[{{Willman} {et~al.}(2005{\natexlab{b}}){Willman}, {Dalcanton},
  {Martinez-Delgado}, {West}, {Blanton}, {Hogg}, {Barentine}, {Brewington},
  {Harvanek}, {Kleinman}, {Krzesinski}, {Long}, {Neilsen}, {Nitta}, \&
  {Snedden}}]{willman05b}
{Willman}, B., {Dalcanton}, J.~J., {Martinez-Delgado}, D., {et~al.}
  2005{\natexlab{b}}, \apjl, 626, L85, \dodoi{10.1086/431760}

\bibitem[{{Woo} {et~al.}(2008){Woo}, {Courteau}, \& {Dekel}}]{woo08}
{Woo}, J., {Courteau}, S., \& {Dekel}, A. 2008, \mnras, 390, 1453,
  \dodoi{10.1111/j.1365-2966.2008.13770.x}

\bibitem[{{Yanny} {et~al.}(2009){Yanny}, {Rockosi}, {Newberg}, {Knapp},
  {Adelman-McCarthy}, {Alcorn}, {Allam}, \& {Allende Prieto}}]{yanny09}
{Yanny}, B., {Rockosi}, C., {Newberg}, H.~J., {et~al.} 2009, \aj, 137, 4377,
  \dodoi{10.1088/0004-6256/137/5/4377}

\bibitem[{{York} {et~al.}(2000){York}, {Adelman}, {Anderson}, {Anderson},
  {Annis}, {Bahcall}, {Bakken}, {Barkhouser}, {Bastian}, {Berman}, \&
  {Boroski}}]{york00}
{York}, D.~G., {Adelman}, J., {Anderson}, Jr., J.~E., {et~al.} 2000, \aj, 120,
  1579, \dodoi{10.1086/301513}

\bibitem[{{Yuan} {et~al.}(2018){Yuan}, {Chang}, {Banerjee}, {Han}, {Kang}, \&
  {Smith}}]{yuan18}
{Yuan}, Z., {Chang}, J., {Banerjee}, P., {et~al.} 2018, \apj, 863, 26,
  \dodoi{10.3847/1538-4357/aacd0d}

\bibitem[{{Yuan} {et~al.}(2020{\natexlab{a}}){Yuan}, {Chang}, {Beers}, \&
  {Huang}}]{yuan20b}
{Yuan}, Z., {Chang}, J., {Beers}, T.~C., \& {Huang}, Y. 2020{\natexlab{a}},
  \apjl, 898, L37, \dodoi{10.3847/2041-8213/aba49f}

\bibitem[{{Yuan} {et~al.}(2019){Yuan}, {Smith}, {Xue}, {Li}, {Liu}, {Wang},
  {Li}, \& {Chang}}]{yuan19}
{Yuan}, Z., {Smith}, M.~C., {Xue}, X.-X., {et~al.} 2019, \apj, 881, 164,
  \dodoi{10.3847/1538-4357/ab2e09}

\bibitem[{{Yuan} {et~al.}(2020{\natexlab{b}}){Yuan}, {Myeong}, {Beers},
  {Evans}, {Lee}, {Banerjee}, {Gudin}, {Hattori}, {Li}, {Matsuno}, {Placco},
  {Smith}, {Whitten}, \& {Zhao}}]{yuan20a}
{Yuan}, Z., {Myeong}, G.~C., {Beers}, T.~C., {et~al.} 2020{\natexlab{b}}, \apj,
  891, 39, \dodoi{10.3847/1538-4357/ab6ef7}

\bibitem[{{Zhao} {et~al.}(2012){Zhao}, {Zhao}, {Chu}, {Jing}, \&
  {Deng}}]{zhao12}
{Zhao}, G., {Zhao}, Y.-H., {Chu}, Y.-Q., {Jing}, Y.-P., \& {Deng}, L.-C. 2012,
  Research in Astronomy and Astrophysics, 12, 723,
  \dodoi{10.1088/1674-4527/12/7/002}

\end{thebibliography}
\bibliographystyle{aasjournal}

\end{document}